\documentclass[a4paper,11pt]{article}
\pdfoutput=1
\usepackage{jheppub}
\usepackage{amsmath}
\newcommand{\ii}{\mathrm{i}}
\newcommand{\rme}{\mathrm{e}}

\newcommand{\Tr}{\mathrm{Tr}\,}
\newcommand{\tr}{\mathrm{tr}\,}
\newcommand{\diag}{\mathrm{diag}\,}

\newcommand{\cW}{{\mathcal{W}}}

\newcommand{\one}{{\rm 1\kern -.9mm l}}

\newcommand{\kfive}{\mathsf{k}_{\text{5d}}}

\newcommand{\be}{\begin{equation}}
\newcommand{\ee}{\end{equation}}
\newcommand{\p}{\partial}

\newdimen\tableauside\tableauside=1.0ex
\newdimen\tableaurule\tableaurule=0.4pt
\newdimen\tableaustep
\def\phantomhrule#1{\hbox{\vbox to0pt{\hrule height\tableaurule
width#1\vss}}}
\def\phantomvrule#1{\vbox{\hbox to0pt{\vrule width\tableaurule
height#1\hss}}}
\def\sqr{\vbox{%
  \phantomhrule\tableaustep
\hbox{\phantomvrule\tableaustep\kern\tableaustep\phantomvrule\tableaustep}%
  \hbox{\vbox{\phantomhrule\tableauside}\kern-\tableaurule}}}
\def\squares#1{\hbox{\count0=#1\noindent\loop\sqr
  \advance\count0 by-1 \ifnum\count0>0\repeat}}
\def\tableau#1{\vcenter{\offinterlineskip
  \tableaustep=\tableauside\advance\tableaustep by-\tableaurule
  \kern\normallineskip\hbox
    {\kern\normallineskip\vbox
      {\gettableau#1 0 }%
     \kern\normallineskip\kern\tableaurule}%
  \kern\normallineskip\kern\tableaurule}}
\def\gettableau#1 {\ifnum#1=0\let\next=\null\else
  \squares{#1}\let\next=\gettableau\fi\next}
\def\XXint#1#2#3{{\setbox0=\hbox{$#1{#2#3}{\int}$}
     \vcenter{\hbox{$#2#3$}}\kern-.5\wd0}}

\usepackage{tikz}
\usetikzlibrary{graphs}
\usetikzlibrary{trees}
\usetikzlibrary{calc}
\usetikzlibrary{decorations.markings}

\tikzstyle{gauge} = [circle, text centered, draw=black, minimum height=1.5cm]
\tikzstyle{flavor} = [rectangle, text centered, draw=black, minimum height=1.5cm,minimum width=1.5cm]

\tikzstyle{gaugeS} = [circle, text centered, draw=black, minimum height=6ex]
\tikzstyle{flavorS} = [rectangle, text centered, draw=black,minimum height=6ex,minimum width=6ex]
\usepackage{siunitx}

\makeatletter
\pgfdeclareshape{barn}{
    \inheritsavedanchors[from={rectangle}]
    \inheritanchorborder[from={rectangle}] 
    \inheritanchor[from={rectangle}]{center}
    \inheritanchor[from={rectangle}]{south}
    \inheritanchor[from={rectangle}]{west}
    \inheritanchor[from={rectangle}]{east}
    \inheritbackgroundpath[from={rectangle}]
   
    \backgroundpath{
    \southwest \pgf@xa=\pgf@x \pgf@ya=\pgf@y
    \northeast \pgf@xb=\pgf@x \pgf@yb=\pgf@y
    \pgf@xc=\pgf@xb \advance\pgf@xc by-\pgf@xa     
    \pgf@yc=\pgf@yb \advance\pgf@yc by-\pgf@ya    
    \advance\pgf@yb by-0.5\pgf@yc
    \pgfpathmoveto{\pgfpoint{\pgf@xa}{\pgf@ya}}
    \pgfpathlineto{\pgfpoint{\pgf@xa}{\pgf@yb}}
    \pgfpatharc{180}{0}{.5\pgf@xc}  
    \pgfpathlineto{\pgfpoint{\pgf@xb}{\pgf@ya}}
    \pgfpathclose
 }
}
\makeatother

\tikzstyle{gaugedflavor} = [barn,draw, text centered, minimum height=1.5cm,minimum width=1.5cm,draw=black]
\tikzstyle{gaugedflavorS} = [barn,draw, text centered, minimum width=6ex,minimum height=6ex,draw=black]

\pgfdeclareshape{cross}{
  \inheritsavedanchors[from={rectangle}]
    \inheritanchorborder[from={rectangle}] 
     \inheritanchor[from={rectangle}]{north}
    \inheritanchor[from={rectangle}]{center}
    \inheritanchor[from={rectangle}]{south}
    \inheritanchor[from={rectangle}]{west}
    \inheritanchor[from={rectangle}]{east}
\inheritbackgroundpath[from={rectangle}]
    \backgroundpath{
    \draw[solid] (.6ex,-.6ex) -- (-.6ex,.6ex);
    \draw[solid] (.6ex,.6ex) -- (-.6ex,-.6ex);
    }
}

\title{\boldmath Surface operators in 5d gauge theories and duality relations}
\author[a]{S.~K.~Ashok,}
\affiliation[a]{Institute of Mathematical Sciences \\
Homi Bhabha National Institute (HBNI)\\
IV Cross Road, C.~I.~T.~Campus, \\
  Taramani, Chennai, 600113  Tamil Nadu, India \\}
\emailAdd{sashok@imsc.res.in} 
   
\author[b,c]{M.~Bill\`o,}
\affiliation[b]{Universit\`a di Torino, Dipartimento di Fisica}

\affiliation[c]{Arnold-Regge Center and I.\,N.\,F.\,N. - sezione di Torino, \\
Via P. Giuria 1, I-10125 Torino, Italy\\}
\emailAdd{billo@to.infn.it}

\author[a]{E.~Dell'Aquila,}
\emailAdd{edellaquila@gmail.com}

\author[b,c]{M.~Frau,}
\emailAdd{frau@to.infn.it}

\author[a]{V.~Gupta,}
\emailAdd{varungupta@imsc.res.in}

\author[b,c]{R.~R.~John,}
\emailAdd{renjan.rajan@to.infn.it}

\author[d,c]{and A.~Lerda\,}
\affiliation[d]{Universit\`a del Piemonte Orientale, Dipartimento di Scienze e Innovazione Tecnologica\\
Viale T. Michel 11, I-15121 Alessandria, Italy\\}
\emailAdd{lerda@to.infn.it}

\abstract{We study half-BPS surface operators in 5d $\mathcal{N}=1$ gauge theories
compactified on a circle. Using localization methods and the twisted chiral ring relations of 
coupled 3d/5d quiver gauge theories, we calculate the twisted chiral superpotential
that governs the infrared properties of these surface operators.
We make a detailed analysis of the localization integrand, and by comparing with the results from 
the twisted chiral ring equations, we obtain constraints on the 3d and 5d Chern-Simons levels so 
that the instanton partition function does not depend on the choice of integration contour. 
For these values of the Chern-Simons couplings, we comment on how the distinct quiver theories 
that realize the same surface operator are related to each other by Aharony-Seiberg dualities.
}

\keywords{Supersymmetric gauge theories, instantons, surface operators, dualities}

\preprint{ARC-17-12 }

\begin{document}
\maketitle
\flushbottom

\section{Introduction and summary}
\label{secn:intro}

Surface operators were first introduced in \cite{Gukov:2006jk, Gukov:2008sn} as half-BPS defects
of codimension two that solve the Kapustin-Witten equations in four-dimensional ${\mathcal N}=4$ 
supersymmetric Yang-Mills theories (see \cite{Gukov:2014gja} for an overview). 
By giving a mass to the adjoint hypermultiplet and flowing 
to the infra-red (IR), these defects naturally lead to surface operators in pure 
${\mathcal N}=2$ gauge theories in four dimensions. 
These surface operators have been extensively studied from many different points of view
\cite{Gaiotto:2009fs, Alday:2009fs, Taki:2009zd, Alday:2010vg,Kozcaz:2010yp,Marshakov:2010fx,
Kozcaz:2010af,Dimofte:2010tz,Maruyoshi:2010iu,
Taki:2010bj,Awata:2010bz,Wyllard:2010vi,Wyllard:2010rp,Kanno:2011fw,Gaiotto:2013sma,
Bullimore:2014awa,Nawata:2014nca,Gomis:2014eya,Frenkel:2015rda,Assel:2016wcr,
Gomis:2016ljm,Pan:2016fbl,Ashok:2017odt,Gorsky:2017hro, Ashok:2017lko,Nekrasov:2017rqy,
Nekrasov:2017gzb}. 

The present paper contains a generalization of our previous work \cite{Ashok:2017lko}, in which 
we studied surface operators in pure Yang-Mills theories with gauge group SU$(N)$ and eight 
supercharges in four and five dimensions, following two approaches. In the first approach, we made use of the 
microscopic description offered by Nekrasov localization \cite{Nekrasov:2002qd,Nekrasov:2003rj}, 
suitably adapted to the case with surface operators 
\cite{Kanno:2011fw, Ashok:2017odt, Gorsky:2017hro, Ashok:2017lko}, and computed the (ramified) 
instanton partition function. 
In the second approach, we considered quiver gauge theories \cite{Gaiotto:2009fs, Gaiotto:2013sma}
in two (or three) dimensions with an additional SU$(N)$ flavour symmetry realized by a
gauge theory in four (or five) dimensions. {From} this standpoint, one deals with combined 
2d/4d (or 3d/5d) systems, whose low-energy effective action is 
encoded in a pair of holomorphic functions: the prepotential, which governs the dynamics in four 
(or five) dimensions, and the twisted chiral superpotential, which describes the massive vacua of 
the quiver theories in two (or three) dimensions. 
Following the general ideas of \cite{Gaiotto:2013sma} and using a careful mapping of parameters, 
in \cite{Ashok:2017lko} we were able to match the twisted superpotential computed
using localization methods with the one obtained by solving the chiral ring equations in 
the quiver theory approach.

In the 2d/4d case there are distinct quiver descriptions for the same surface operator 
\cite{Gorsky:2017hro, Ashok:2017lko} in which the corresponding 2d theories are related to each other
by Seiberg-like dualities \cite{Seiberg:1994pq,Benini:2014mia,Closset:2015rna}. 
{From} the localization point of view, these distinct ultra-violet (UV) descriptions correspond to 
different choices of the integration contours along which one computes the integral 
over the (ramified) instanton moduli space to obtain the Nekrasov partition function. 
When these theories are lifted to 3d/5d systems, some novel features arise. Indeed, as we have shown 
in \cite{Ashok:2017lko}, suitable Chern-Simons terms in three dimensions are needed in order to ensure 
the equality of the twisted superpotentials in dual descriptions. 
This is not too surprising since the 3d quiver theories include bi-fundamental matter multiplets that 
are rendered massive by twisted masses. When one integrates out these massive chiral fields, 
one generates effective Chern-Simons interactions. Furthermore, since dual pairs in three dimensions 
are related by Aharony-Seiberg dualities \cite{Seiberg:1994pq,Aharony:1997gp, Aharony:2014uya} 
which typically act on the Chern-Simons levels, we expect that the 
Chern-Simons couplings of two different quiver theories describing the same surface operator
must be related in a precise manner.
In \cite{Ashok:2017lko} a few examples were worked out to highlight this phenomenon. We showed 
that the effective twisted chiral superpotential matched only for particular values of the 3d 
Chern-Simons levels in the dual pairs. In this work, we perform a complete and systematic analysis of 
coupled 3d/5d theories that have an interpretation as supersymmetric surface operators 
in ${\mathcal N}=1$ gauge theories in five dimensions, allowing for both 3d as well as 5d Chern-Simons
interactions, and provide a general description of the duality relations. 

We now give an overview of this paper. In Section~\ref{secn:5dCS}, we review the localization 
analysis of the 5d $\mathcal{N}=1$ super Yang-Mills theory compactified on a circle
and present its instanton partition function, mainly following \cite{Tachikawa:2004ur} 
(see also \cite{Kim:2012gu,Bergman:2013ala,Bergman:2013aca,Taki:2013vka,Taki:2014pba,
Hwang:2014uwa}). 
However, instead of directly working with the Young tableaux formulation, we work with the contour
integral formulation.

In Section~\ref{localization}, we study the 5d SU($N$) theories in the presence 
of surface operators, which we treat as monodromy defects \cite{Gukov:2006jk, Gukov:2008sn} labeled 
by the partitions of $N$ of length $M$. For any given partition we present the ramified 
instanton partition function that is obtained by a suitable $\mathbb{Z}_M$ orbifold projection 
on the instanton moduli space of the theory without defects
\cite{Alday:2009fs, Kanno:2011fw, Ashok:2017odt}. The integrand of this ramified instanton partition
function has the same set of poles as the one presented in \cite{Ashok:2017lko} but it has additional exponential factors that depend on $M$ new parameters, which we denote $\mathsf{m}_I$,
whose sum plays the role of the Chern-Simons 
coupling $\kfive$ of the five dimensional SU$(N)$ gauge theory.
To obtain explicit results, one must specify the integration contours in the instanton partition function, 
which can be conveniently classified by a Jeffrey-Kirwan reference vector \cite{JK1995} 
(see \cite{Gorsky:2017hro,Ashok:2017lko} for details). Here we present two choices 
which are complementary to each other and are simple extensions of those discussed in the pure 
5d theory. For these two choices 
we compute the twisted chiral superpotential by explicitly evaluating the residues over the poles 
selected by the integration contours.

In Section~\ref{secn:3d5d}, we go on to study surface operators as coupled 3d/5d  systems and identify 
two quiver descriptions with $(M-1)$ 3d gauge nodes and an SU$(N)$ flavour node that is gauged in
five dimensions, which are dual to each other. 
The identification proceeds as follows: for a given 3d/5d quiver theory, we solve the 
twisted chiral ring equations about a particular classical vacuum as a power series expansion in the 
strong coupling scales of the quiver theory. 
Then, we show that there is a one-to-one map between the choice of classical vacuum and the 
choice of discrete data that label a Gukov-Witten defect. In particular, the strong coupling 
scales of the 3d/5d quiver 
are mapped on to the Nekrasov instanton counting parameters, while the Chern-Simons levels of the 3d 
nodes of the quiver theory are related to the first $(M-1)$ parameters $\mathsf{m}_I$ of the localization
calculation. However, the precise map depends on the choice of the contour prescription. In fact, 
with one prescription, these parameters are related to the Chern-Simons levels, but with the other 
they are related to the negative of the Chern-Simons levels.

In Section~\ref{secn:AharonySeiberg}, we revisit the conditions under which the two contour prescriptions
yield equal results and interpret them as Aharony-Seiberg dualities \cite{Seiberg:1994pq, Aharony:1997gp}
between pairs of quiver theories. In this correspondence we find that the 3d Chern-Simons 
levels are integral or half-integral, depending on the ranks of the 3d/5d quiver. These constraints 
coincide with those derived 
in \cite{Niemi:1983rq,Redlich:1983kn,Redlich:1983dv} by requiring the absence of a 
parity anomaly. Further, we find that the bounds on the 3d Chern Simons levels are the same as the ones 
obtained in \cite{Benini:2011mf} for what are called maximally chiral theories. While Aharony-Seiberg dual 
pairs exist for other types of 3d quivers also, it is only for the maximally chiral ones that the 
twisted masses (induced by the 5d Coulomb vacuum expectation values) completely lift the 3d 
Coulomb moduli space and render the 3d theory completely massive. This is consistent with the general 
analysis of \cite{Gaiotto:2009fs} where it was shown that only the 2d (or 3d) massive theories 
can be embedded as surface operators in four (or five) dimensions. We therefore 
conclude that it is precisely such maximally chiral theories that have avatars as surface 
operators in 5d theories. 

Finally, we collect some technical material in the appendices.

\section{5d gauge theories}
\label{secn:5dCS}
In this section we describe the derivation of the instanton partition function for a 
gauge theory with a Chern-Simons term in five dimensions, following the analysis 
of \cite{Tachikawa:2004ur} that relies on the use of localization methods. This partition function
has already been extensively studied in the literature (see for example 
\cite{Kim:2012gu,Bergman:2013ala,Bergman:2013aca,Taki:2013vka,Taki:2014pba,Hwang:2014uwa}) 
but we review it here to set the stage for the analysis in the following sections. We then consider 
the resolvent of the 5d theory from the point of view of 
the Seiberg-Witten curve and establish a connection with the localization methods.

\subsection{Localization}
\label{pure5dsection}
Let us consider an $\mathcal{N}=1$ SU($N$) gauge theory 
defined on $\mathbb{R}^4\times S^1$, and denote by $\beta$ the length of the circumference 
$S^1$ and by $\kfive$ the coefficient of the Chern-Simons term. 
We study this theory in a generic point in the Coulomb branch parameterized by 
the vacuum expectation values $a_u$ (with $u=1,\cdots, N$) of
the adjoint scalar field $\Phi$ in the vector multiplet, that satisfy the 
SU($N$) tracelessness condition
\begin{equation}
\sum_{u=1}^Na_u=0
\label{SUN}
\end{equation}
but are otherwise arbitrary. Being at a generic point of the Coulomb branch, 
according to the analysis of \cite{Intriligator:1997pq}, we must take
\begin{equation}
\kfive \in\mathbb{Z}\qquad\mbox{and}\qquad
 \left\vert \,\kfive\, \right\vert \leq N~.
 \label{IMSbound}
\end{equation}
The integrality constraint is a consequence of analyzing the non-compact 5d theory on the Coulomb branch 
and imposing gauge invariance of the resulting cubic prepotential, while the bound on $\kfive$
comes from requiring that the 5d gauge theory has an interacting UV fixed point on the entire
Coulomb branch\,\footnote{See also the recent work \cite{Jefferson:2017ahm,Jefferson:2018irk} 
in which the results 
of \cite{Intriligator:1997pq} have been generalized by requiring that only a subspace of the 
Coulomb moduli space be physical. It would be interesting to investigate if the localization approach 
we are describing can be applied also to this case where novel massless degrees of freedom occur, 
but this is beyond the scope of this paper.}.

After deforming $\mathbb{R}^4$ by an $\Omega$-background \cite{Nekrasov:2002qd,Nekrasov:2003rj} 
parametrized by $\epsilon_1$ and $\epsilon_2$, we use localization methods to compute the 
partition function in the instanton sector.
This can be written as
\begin{equation}
Z_{\text{inst}} = 1+\sum_{k=1}^{\infty} \frac{(-q)^k}{k!} \int_{\mathcal C}\prod_{\sigma=1}^k \bigg(
\beta\,\frac{d\chi_\sigma}{2\pi\ii}\bigg) \,z_k(\chi_\sigma)
\label{ZY5d}
\end{equation}
where
\begin{equation}
\begin{aligned}
\label{5dCSintegrand}
z_k(\chi_\sigma) &= \rme^{-\beta\, \kfive \sum_{\sigma}\chi_\sigma}\!
\prod_{\sigma,\tau=1}^k\left[\frac{g\big(\chi_{\sigma}-
\chi_{\tau}+\epsilon_1+\epsilon_2\big)}{g\big(\chi_{\sigma}-\chi_{\tau}+\epsilon_1\big)
\,g\big(\chi_{\sigma}-\chi_{\tau}+\epsilon_2\big)} \right]
\prod_{\substack{\sigma,\tau=1\\ \sigma\not=\tau}}^k \!g\big(\chi_{\sigma}-\chi_{\tau}\big)
\\
&\hspace{2cm}\times\prod_{\sigma=1}^k\prod_{u=1}^N \left[\frac{1}{g\big(\chi_\sigma-a_u+
\frac{\epsilon_1+\epsilon_2}{2}\big)\,g\big(-\chi_\sigma+a_u+\frac{\epsilon_1
+\epsilon_2}{2}\big)}\right]~.
\end{aligned}
\end{equation}
and \cite{Nekrasov:2002qd, Nekrasov:2003rj, Hollowood:2003cv,Tachikawa:2004ur}
\begin{equation}
g(x)=2\sinh\left(\frac{\beta \,x}{2}\right)~.
\end{equation}
We observe that the Chern-Simons coefficient $\kfive$ only appears 
in the exponent of the prefactor in (\ref{5dCSintegrand}). The instanton counting 
parameter $q$ is given by
\begin{equation}
q = (-1)^N(\beta \Lambda)^{2N}
\end{equation}
where $\Lambda$ is the (complexified) strong-coupling scale. 
It is easy to check that in the limit $\beta\to 0$ the above expressions reduce to those 
appropriate for a pure $\mathcal{N}=2$ super Yang-Mills theory in four 
dimensions with SU($N$) gauge group and dynamically generated scale $\Lambda$.

The integral in (\ref{ZY5d}) is performed on a closed contour $\mathcal{C}$ in the complex 
$\chi_\sigma$-plane which has to be suitably chosen in such a way that it surrounds a finite number 
of singularities of the integrand function.
If we make the standard choice for the imaginary part of the $\Omega$-background parameters, namely
\begin{equation}
1\gg \mathrm{Im}\, \epsilon_1\gg \mathrm{Im}\, 
\epsilon_2 > 0~,
\label{e1e2ImaginaryPart}
\end{equation}
and take $a_u$ to be real for simplicity, the poles of (\ref{5dCSintegrand}) lie 
either in the upper or in the lower-half 
complex $\chi_\sigma$-plane, and can be put in correspondence with an $N$-array of Young 
tableaux $\{Y_u \}$ such that the total number of boxes is equal to the instanton number $k$ 
\cite{Nekrasov:2002qd}\,\footnote{For further details we refer for example 
to \cite{Billo:2012st,Hwang:2014uwa}.}. 
More precisely, the poles of (\ref{5dCSintegrand}) are located at
\begin{equation}
\chi_\sigma =  a_u \pm \Big(i - \frac{1}{2}\Big)\epsilon_1 \pm 
\Big(j-\frac{1}{2}\Big)\epsilon_2 + \frac{2\pi\ii}{\beta}\,n
\label{poles}
\end{equation}
where $(i,j)$ run over the rows and columns of the Young tableau $Y_u$ and the last 
term, proportional to the integer $n$, is due to the periodicity of the $\sinh$-function of a 
complex variable. Notice that the Chern-Simons coupling $\kfive$ 
does not affect the location of the poles and it only adds additional multiplicative factors 
to the residue at each pole. 

\begin{figure}[ht]
\begin{tikzpicture}[decoration={markings,mark=at position 0.4 with {\draw[solid] (1ex,-1ex) 
-- (0ex,0ex);
\draw[solid] (1ex,1ex) -- (0ex,0ex);}}]                
\draw[thin,->] (-7,0) -- (7.3,0) node[anchor=80] {$\mathrm{Re}(\chi_{\sigma})$ };
\draw[thin,->] (0,-4) -- (0,4) node[anchor=south west] {$\mathrm{Im}(\chi_{\sigma})$};
\foreach \y in {2,3}
  \draw (2pt,\y cm) -- (-2pt,\y cm) node[anchor=east] {$\frac{\y \pi}{\beta}$};
\draw (2pt,-3) -- (-2pt,-3) node[anchor=east] {$-\frac{3\pi}{\beta}$};     
\draw (2pt,-2) -- (-2pt,-2) node[anchor=east] {$-\frac{2\pi}{\beta}$};  
\draw (2pt,-1) -- (-2pt,-1) node[anchor=east] {$-\frac{\pi}{\beta}$};  
\foreach \x in {2,4,-3}
\draw[gray,thick,dotted] (\x,-4)--(\x,4);
\draw[very thick,blue!80] (-5.5,1) --(5.5,1)[postaction={decorate}] node[anchor=-60] {{\large 
$\mathcal{C}^{(\sigma)}_+$} };
\draw[very thick,blue!80,dashed] (5.5,1) --(6.1,1); 
\draw[very thick,blue!80,dashed] (6.1,1) --(6.1,0)[postaction={decorate}]; 
\draw[very thick,blue!80,dashed] (6.1,0) --(5.5,0); 
\draw[very thick,blue!80] (5.5,0) --(-5.5,0)[postaction={decorate}];  
\draw[very thick,blue!80,dashed] (-5.5,0) --(-6.1,0); 
\draw[very thick,blue!80,dashed] (-6.1,0) --(-6.1,1)[postaction={decorate}]; 
\draw[very thick,blue!80,dashed] (-6.1,1) --(-5.5,1); 
\foreach \i in {-4,-2,2}
   \draw ($(2,0.3+\i)$) node[cross,black!85,very thick]{}
   ($(4,0.3+\i)$) node[cross,black!85,very thick]{} 
   ($(-3,0.3+\i)$) node[cross,black!85,very thick]{};
\draw ($(2,0.3)$) node[cross,blue!80,very thick]{}
   ($(4,0.3)$) node[cross,blue!80,very thick]{} 
   ($(-3,0.3)$) node[cross,blue!80,very thick]{};
\foreach \i in {-2,2,4}
   \draw ($(2,-0.3+\i)$) node[cross,black!55,very thick]{}
   ($(4,-0.3+\i)$) node[cross,black!55,very thick]{} 
   ($(-3,-0.3+\i)$) node[cross,black!55,very thick]{};
 \draw ($(2,-0.3)$) node[cross,red,very thick]{}
   ($(4,-0.3)$) node[cross,red,very thick]{} 
   ($(-3,-0.3)$) node[cross,red,very thick]{};
\end{tikzpicture}
\caption{For each integration variable $\chi_\sigma$, the fundamental domain is the 
region $-\infty < \mathrm{Re}\chi_\sigma < \infty$ and $-\frac{\pi}{\beta} < \mathrm{Im}\chi_\sigma 
< \frac{\pi}{\beta}$. The poles in the fundamental domain are shown in colour. 
The contour ${\mathcal C}^{(\sigma)}_+$ selects those poles in the fundamental domain 
that are in the upper half plane. In this picture we have explicitly shown the 1-instanton case 
for the SU(3) gauge theory at $k=1$.}
\label{contourC+}
\end{figure}
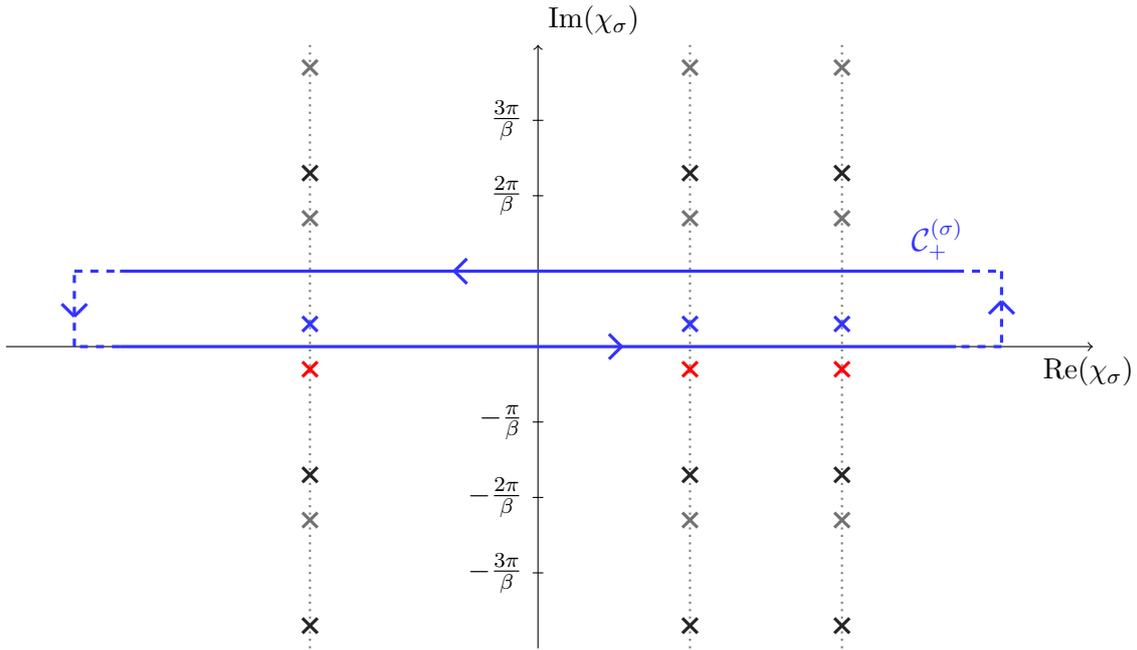

\begin{figure}[ht]
\begin{tikzpicture}[decoration={markings,mark=at position 0.4 with {\draw[solid] (1ex,-1ex) 
-- (0ex,0ex);
                \draw[solid] (1ex,1ex) -- (0ex,0ex);}}]
\draw[thin,->] (-7,0) -- (7.3,0) node[anchor=80] {$\mathrm{Re}(\chi_{\sigma})$ };
\draw[thin,->] (0,-4) -- (0,4) node[anchor=south west] {$\mathrm{Im}(\chi_{\sigma})$};
\foreach \y in {2,3}
    \draw (2pt,\y cm) -- (-2pt,\y cm) node[anchor=east] {$\frac{\y \pi}{\beta}$};
     \draw (2pt,-2 cm) -- (-2pt,-2 cm) node[anchor=east] {$-\frac{2 \pi}{\beta}$};
      \draw (2pt,-3 cm) -- (-2pt,-3 cm) node[anchor=east] {$-\frac{3 \pi}{\beta}$};
\draw (2pt,1) -- (-2pt,1) node[anchor=east] {$\frac{\pi}{\beta}$};  

\foreach \x in {2,4,-3}
\draw[gray,thick,dotted] (\x,-4)--(\x,4); 

\draw[very thick,red] (-5.5,-1) --(5.5,-1)[postaction={decorate}] ;
\draw[very thick,red,dashed] (5.5,-1) --(6.1,-1); 
\draw[very thick,red,dashed] (6.1,-1) --(6.1,0)[postaction={decorate}]; 
\draw[very thick,red,dashed] (6.1,0) --(5.5,0); 
\draw[very thick,red] (5.5,0) node[anchor=-60] {{\large $\mathcal{C}^{(\sigma)}_-$}}--(-5.5,0)[postaction={decorate}];  
\draw[very thick,red,dashed] (-5.5,0) --(-6.1,0); 
\draw[very thick,red,dashed] (-6.1,0) --(-6.1,-1)[postaction={decorate}]; 
\draw[very thick,red,dashed] (-6.1,-1) --(-5.5,-1); 

\foreach \i in {-4,-2,2}
   \draw ($(2,0.3+\i)$) node[cross,black!85,very thick]{}
   ($(4,0.3+\i)$) node[cross,black!85,very thick]{} 
   ($(-3,0.3+\i)$) node[cross,black!85,very thick]{};
\draw ($(2,0.3)$) node[cross,blue!80,very thick]{}
   ($(4,0.3)$) node[cross,blue!80,very thick]{} 
   ($(-3,0.3)$) node[cross,blue!80,very thick]{};

\foreach \i in {-2,2,4}
   \draw ($(2,-0.3+\i)$) node[cross,black!55,very thick]{}
   ($(4,-0.3+\i)$) node[cross,black!55,very thick]{} 
   ($(-3,-0.3+\i)$) node[cross,black!55,very thick]{};
 \draw ($(2,-0.3)$) node[cross,red,very thick]{}
   ($(4,-0.3)$) node[cross,red,very thick]{} 
   ($(-3,-0.3)$) node[cross,red,very thick]{};
\end{tikzpicture}
\caption{For each integration variable $\chi_\sigma$, the contour ${\mathcal C}^{(\sigma)}_{-}$ 
selects those poles in the 
fundamental domain that are in the lower half plane. Once again, the poles that are shown 
in this picture are those for the SU$(3)$ gauge theory at $k=1$.}
\label{contourC-}
\end{figure}
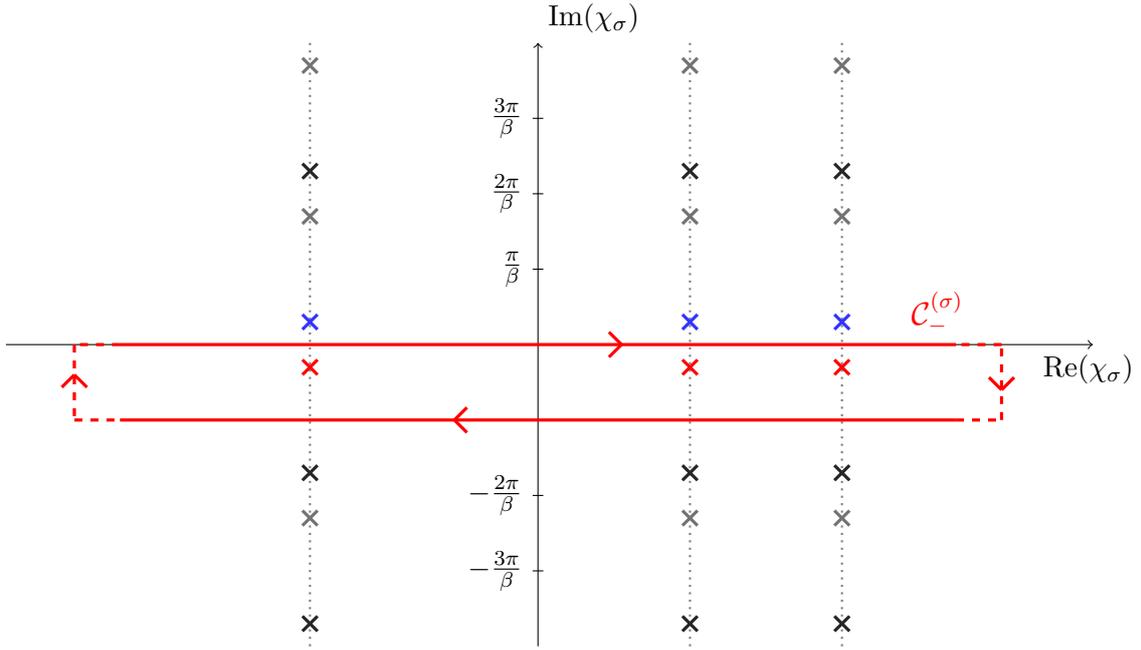

When we restrict to a fundamental domain by setting $n=0$ in (\ref{poles}), 
we have only two sets of poles\,\footnote{We observe that these 
two sets of poles are the same 
that are considered in the corresponding calculation in four dimensions \cite{Nekrasov:2003rj}.}:
those that are just above the real axis and those that are just below it. Each of these two sets leads 
precisely to the results of 
\cite{Tachikawa:2004ur,Bergman:2013ala,Bergman:2013aca,Taki:2013vka,Taki:2014pba,Hwang:2014uwa}.
The poles in the region
\begin{equation}
0 <\text{Im}\, \chi_\sigma  < \frac{\pi}{\beta}
\end{equation}
are selected by the contour $\mathcal{C}^{(\sigma)}_+$ as in
Fig.~\ref{contourC+} for the SU(3) theory at $k=1$.
Instead, the poles in the region
\begin{equation}
 -\frac{\pi}{\beta}<\text{Im}\, \chi_\sigma<0 
\end{equation}
are selected by the contour ${\mathcal C}^{(\sigma)}_-$ as in 
Fig.~\ref{contourC-}, again for the SU(3) theory at $k=1$.

In both cases, the contours extend all the way to infinity along the horizontal direction, 
since the positions of the poles can have arbitrary real parts because the vacuum expectation 
values $a_u$ are only subject to the condition (\ref{SUN}) but
are otherwise arbitrary. Another issue is the fact that the two integration contours 
${\mathcal C}^{(\sigma)}_{\pm}$ may lead to different results.
To illustrate the main ideas, it suffices to consider the 1-instanton term of the partition function, namely
\begin{equation}
\begin{aligned}
\label{Partition_function_TachikawaFactor}
Z_{1-\text{inst}} &= 
               -q\,\frac{g\big(\epsilon_1+\epsilon_2\big)}{g\big(\epsilon_1\big)\,g\big(\epsilon_2\big)}\!
               \int_{\mathcal{C}}\! \Big(\beta\,\frac{d\chi}{2\pi\ii}\Big)\, \rme^{- \beta\, \kfive \,\chi}
               \prod_{u=1}^N\frac{1}{g\big(\chi-a_u+\frac{\epsilon_1+\epsilon_2}{2}\big)\,
               g\big(\!-\chi+a_u+\frac{\epsilon_1+\epsilon_2}{2}\big)} ~.
\end{aligned}
\end{equation}
We find it convenient to perform the following change of variables 
\begin{equation}
\label{bigvars}
 \chi = \frac{1}{ \beta } \log X\, ,\quad a_u=\frac{1}{\beta}\log A_u,\quad \epsilon_1 
 = \frac{1}{ \beta } \log E_1 \,,\quad \text{and} \,\,\, \epsilon_2 = \frac{1}{ \beta } \log E_2 ~,
\end{equation}
and rewrite  (\ref{Partition_function_TachikawaFactor}) as
\begin{align}
\label{Partition_function_bigvars}
Z_{1-\text{inst}} =&- q\,\frac{E_1E_2-1}{(E_1-1)(E_2-1)}
\int_{\mathcal{C}} \frac{dX}{2\pi\ii}\,X^{N-1-\kfive}\,\prod_{u=1}^N
\frac{\sqrt{E_1E_2}}{\big(X\sqrt{E_1E_2}-A_u\big)
\big(A_u\sqrt{E_1E_2}-X\big)}~.
\end{align}
Here we have exploited  the tracelessness condition (\ref{SUN}), which in the new variables becomes
\begin{equation}
\prod_{u=1}^N A_u =1~.
\label{SUNA}
\end{equation}
Under the map (\ref{bigvars}) the regions $0 <\text{Im}\, \chi< \frac{\pi}{\beta}$ 
and $ -\frac{\pi}{\beta} <\text{Im}\, \chi < 0$ transform, respectively, onto the regions 
$\text{Im}\, X > 0$
and $\text{Im}\, X < 0$, and thus the fundamental domain of the $\chi$-plane is mapped 
onto the entire $X$-plane. Furthermore,
the original integration contours $\mathcal{C}_\pm$
are mapped to the (infinite) semi-circles as shown in Fig.~\ref{contoursXplane1} 
and Fig.~\ref{contoursXplane2}. Therefore, choosing the contour ${\mathcal C}_{+}$ or 
${\mathcal C}_{-}$ corresponds to choosing the poles of the integrand of (\ref{Partition_function_bigvars}),
respectively in the upper- or in lower-half complex $X$-plane, that is
\begin{equation}
\begin{aligned}
X&=\,A_u \sqrt{E_1\,E_2}\quad\quad{\mbox{for}}\quad \mathcal{C}_+~,\\
X&=\,\frac{A_u}{\sqrt{E_1\,E_2}}\quad\qquad~{\mbox{for}}\quad \mathcal{C}_-~.
\end{aligned}
\label{polesX}
\end{equation}
\begin{center}
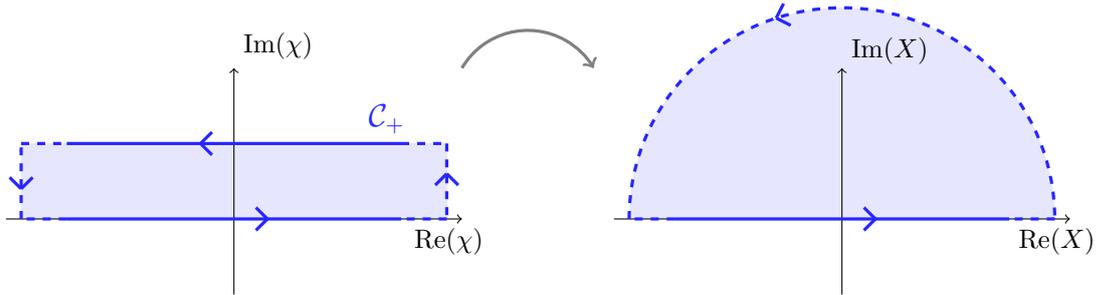
\begin{figure}[ht]
\begin{tikzpicture}[decoration={markings,mark=at position 0.4 with {\draw[solid] (1ex,-1ex) 
-- (0ex,0ex);
                \draw[solid] (1ex,1ex) -- (0ex,0ex);}}]
\small
\fill[blue!10] (-6.8,0) -- (-1.2,0) -- (-1.2,1)--(-6.8,1);

\draw[thin,->] (-7,0) -- (-1,0) node[anchor=60] {$\mathrm{Re}(\chi)$ };
\draw[thin,->] (-4,-1) -- (-4,2) node[anchor=south west] {$\mathrm{Im}(\chi)$};

\draw[very thick,blue!85] (-6.2,1) --(-1.8,1)[postaction={decorate}] node[anchor=-60] {{\large $\mathcal{C}_+$} };
\draw[very thick,blue!85,dashed] (-1.8,1) --(-1.2,1); 
\draw[very thick,blue!85,dashed] (-1.2,1) --(-1.2,0)[postaction={decorate}]; 
\draw[very thick,blue!85,dashed] (-1.2,0) --(-1.8,0); 
\draw[very thick,blue!85] (-1.8,0) --(-6.2,0)[postaction={decorate}];  
\draw[very thick,blue!85,dashed] (-6.2,0) --(-6.8,0); 
\draw[very thick,blue!85,dashed] (-6.8,0) --(-6.8,1)[postaction={decorate}]; 
\draw[very thick,blue!85,dashed] (-6.8,1) --(-6.2,1); 

\fill[blue!10] (1.2,0) arc (180:0:2.8cm);

\draw[thin,->] (1,0) -- (7,0) node[anchor=60] {$\mathrm{Re}(X)$ };
\draw[thin,->] (4,-1) -- (4,2) node[anchor=-160] {$\mathrm{Im}(X)$};

\draw[very thick,blue!85,dashed] (6.8,0) --(6.2,0); 
\draw[very thick,blue!85] (6.2,0) --(1.8,0)[postaction={decorate}];  
\draw[very thick,blue!85,dashed] (1.8,0) --(1.2,0); 
\draw[very thick,blue!85,dashed] (1.2,0) arc (180:0:2.8cm)[postaction={decorate}];
\draw[very thick, gray,->] (-1,2) arc (150:30:1cm);
\end{tikzpicture}
\caption{Map of the contour ${\mathcal C}_+$ from the $\chi$-plane to the $X$-plane.}
\label{contoursXplane1}
\end{figure}
\end{center}
\begin{center}
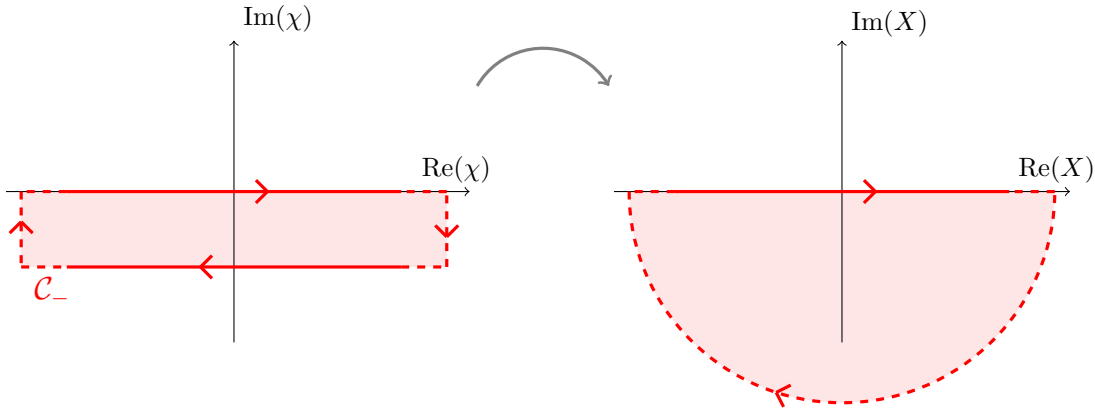
\begin{figure}[ht]
\begin{tikzpicture}[decoration={markings,mark=at position 0.4 with {\draw[solid] (1ex,-1ex) -- (0ex,0ex);
                \draw[solid] (1ex,1ex) -- (0ex,0ex);}}]
\small
\fill[red!10] (-6.8,0) -- (-1.2,0) -- (-1.2,-1)--(-6.8,-1);

\draw[thin,->] (-7,0) -- (-0.9,0) node[anchor=-60] {$\mathrm{Re}(\chi)$ };
\draw[thin,->] (-4,-2) -- (-4,2) node[anchor=south west] {$\mathrm{Im}(\chi)$};

\draw[very thick,red] (-6.2,-1)node[anchor=60] {{\large $\mathcal{C}_-$} } --(-1.8,-1)[postaction={decorate}] ;
\draw[very thick,red,dashed] (-1.8,-1) --(-1.2,-1); 
\draw[very thick,red,dashed] (-1.2,-1) --(-1.2,0)[postaction={decorate}]; 
\draw[very thick,red,dashed] (-1.2,0) --(-1.8,0); 
\draw[very thick,red] (-1.8,0) --(-6.2,0)[postaction={decorate}];  
\draw[very thick,red,dashed] (-6.2,0) --(-6.8,0); 
\draw[very thick,red,dashed] (-6.8,0) --(-6.8,-1)[postaction={decorate}]; 
\draw[very thick,red,dashed] (-6.8,-1) --(-6.2,-1); 

\fill[red!10] (1.2,0) arc (180:360:2.8cm);

\draw[thin,->] (1,0) -- (7,0) node[anchor=-60] {$\mathrm{Re}(X)$ };
\draw[thin,->] (4,-2) -- (4,2) node[anchor=-160] {$\mathrm{Im}(X)$};

\draw[very thick,red,dashed] (6.8,0) --(6.2,0); 
\draw[very thick,red] (6.2,0) --(1.8,0)[postaction={decorate}];  
\draw[very thick,red,dashed] (1.8,0) --(1.2,0); 
\draw[very thick,red,dashed] (1.2,0) arc (180:360:2.8cm)[postaction={decorate}];

\draw[very thick, gray,->] (-0.8,1.4) arc (150:30:1cm);
\end{tikzpicture}
\caption{Map of the contour ${\mathcal C}_-$ from the $\chi$-plane to the $X$-plane.}
\label{contoursXplane2}
\end{figure}
\end{center}

In this formulation it is evident that the constraint $\kfive \in \mathbb{Z}$ implies the absence
of branch cuts in $X$; furthermore if
\begin{equation}
\label{cond2k5d}
 \left\vert \,\kfive\, \right\vert \leq N-1~,
\end{equation}
one can easily see that the instanton partition function receives contributions only from
from the physical poles (\ref{polesX}) or, equivalently, it has no contributions 
from $X=0$ and $X=\infty$. Therefore, when the condition (\ref{cond2k5d}) is satisfied, the two 
different integration prescriptions lead to the same result for the partition function since 
the contours $\mathcal{C}_+$ and $\mathcal{C}_-$
can be smoothly deformed into each other.

Notice that in the original $\chi$-variable, imposing the condition $\kfive \in \mathbb{Z}$ is equivalent to requiring that the integrand function be periodic with period 
$\frac{2\pi}{\beta}$. When this is the case, the two contours $\mathcal{C}_\pm$ are 
equivalent to each other provided the contributions of the vertical segments 
at $\mathrm{Re}\,\chi=\pm\infty$ vanish. This happens precisely 
when (\ref{cond2k5d}) is satisfied. It is interesting to observe that when $\vert\kfive\vert=N$, 
the two contours $\mathcal{C}_\pm$
are not equivalent to each other due to the presence of a residue either at $X=0$ for $\kfive=N$, or
at $X=\infty$ for $\kfive=-N$.
However, these residues are independent of $A_u$ and $N$ since the singularities are simple
poles. They are related to the partition function of an ``\,SU(1)\,'' theory at level $\pm1$ 
\cite{Bergman:2013ala,Bergman:2013aca}\,\footnote{This can be easily 
seen by taking (\ref{Partition_function_bigvars})
and (\ref{SUNA}) for $N=1$ and $\kfive=\pm1$.}, and thus
can be interpreted as the contribution of a continuum in the Coulomb branch 
which has to be suitably 
taken into account and decoupled in order to properly define the SU($N$) theory at $\kfive=\pm N$
\cite{Bergman:2013ala,Bergman:2013aca,Taki:2013vka,Taki:2014pba,Hwang:2014uwa}.
In this way we recover via the contour analysis that the five dimensional Chern-Simons coupling 
satisfies the constraint obtained by \cite{Intriligator:1997pq}. For simplicity, in the following we will restrict ourselves to $\kfive$ as in (\ref{cond2k5d}).

\subsection{Seiberg-Witten curve and resolvent}
\label{sectionResolvent}

We now review the Seiberg-Witten geometry \cite{Seiberg:1994rs, Nekrasov:1996cz} of an SU($N$) gauge theory on 
$\mathbb{R}^4\times S^1$ and propose an all-order expression for the resolvent which we shall 
verify using explicit localization methods. The Seiberg-Witten curve 
can be derived from different approaches. One way is to study M-theory 
on the resolution of non-compact toric Calabi-Yau spaces, 
the so-called $Y^{p,q}$ manifolds, which give rise to SU$(p)$ gauge theories with $\kfive=q$ 
\cite{Hollowood:2003cv, Hanany:2005hq, Brini:2008rh}\footnote{One can also study the gauge theories 
using 5-brane webs that are dual to the toric Calabi-Yau \cite{Aharony:1997ju,Aharony:1997bh}.}. 
The corresponding Seiberg-Witten curve 
is identified with the mirror curve of the local (toric) Calabi-Yau space \cite{Hori:2000kt, Hollowood:2003cv}. 
In most of the literature, the $Y^{p,q}$ spaces are defined with $0<q<p$ and thus only positive 
values of the Chern-Simons level are considered\,\footnote{The boundary values 
$q=0$ and $q=p$ are discussed in \cite{Hanany:2005hq, Brini:2008rh}. }. However, as we will 
see momentarily, the form of the resulting Seiberg-Witten curve is also valid for negative values 
of $\kfive$, although there are interesting subtleties that arise while comparing 
with localization analysis. An alternative approach is to use the NS5-D4 brane set up \cite{Witten:1997sc}
to engineer the classical gauge theory and study its M-theory lift 
\cite{Brandhuber:1997ua}. 
Both approaches give identical results and the Seiberg-Witten curve for a 5d
SU$(N)$ gauge theory with Chern-Simons level $\kfive$ takes the following 
form
\begin{equation}
\label{SWcurvek5d}
Y^2 = {P}_N^2(Z) - 4(\beta \Lambda)^{2N} Z^{-\kfive}
\end{equation}
where
\begin{equation}
\label{PNZexp}
P_N(Z) = Z^{-\frac{N}{2}}\Big(Z^N + \sum_{i=1}^{N-1} (-1)^i \,Z^{N-i}\,
U_i(\kfive)+ (-1)^N \Big) ~.
\end{equation}
Here $U_i(\kfive)$ are the gauge invariant coordinates on the Coulomb branch of the 5d
theory. They are the quantum completion of the classical symmetric polynomials
\begin{equation}
U_i^{\mathrm{class}}= \sum_{u_1\neq u_2\cdots\neq u_i=1}^N
A_{u_1}A_{u_2}\cdots A_{u_i}
\label{Ucl}
\end{equation}
in the vacuum expectation values $A_u$ subjected to the tracelessness condition (\ref{SUNA}), and
explicitly depend on the Chern-Simons level. Notice that if
we use (\ref{Ucl}) in (\ref{PNZexp}), we simply obtain
\begin{equation}
P_N^{\,\mathrm{class}}(Z)=Z^{-\frac{N}{2}}\,\prod_{u=1}^N(Z-A_u)~,
\label{PNclass}
\end{equation}
which is the expected classical expression for $P_N$.
If we now impose the condition that the right hand side of (\ref{SWcurvek5d}) is a doubly monic 
Laurent polynomial in $Z$, it follows that the absolute value of the Chern-Simons level $|\kfive|$ 
has to be an integer smaller than $N$. We thereby recover the constraint
(\ref{cond2k5d}) from the geometry of the Seiberg-Witten curve. While the curve (\ref{SWcurvek5d}) 
was derived for positive values of $\kfive$, it is easy to realize that it holds for negative values 
as well. Indeed, from the brane-web construction, one can show that changing the sign of the Chern-Simons 
coupling amounts to a $\pi$-rotation of the brane configuration. In our explicit realization this corresponds to
\begin{equation}
Z\rightarrow \frac{1}{Z} \qquad\mbox{and}\qquad A_u\rightarrow \frac{1}{A_u}~.
\end{equation}
Let us start from the curve (\ref{SWcurvek5d}) with a positive $\kfive$ and perform the
above map. This yields
\begin{equation}
Y^2 = Z^{-N}\Big(Z^N + \sum_{i=1}^{N-1} (-1)^i \,Z^{N-i}\,\widetilde{U}_{N-i}(\kfive) + (-1)^N \Big)^2 
- 4(\beta\Lambda)^{2N} Z^{\kfive}
\label{SW1}
\end{equation}
where $\widetilde{U}_i$ is obtained from $U_i$ under the inversion of $A_u$. By setting
\begin{equation}
\widetilde{U}_{N-i}(\kfive) =U_i(-\kfive)~,
\label{UUtilde}
\end{equation}
we can rewrite (\ref{SW1}) as
\begin{equation}
Y^2 = Z^{-N}\Big(Z^N + \sum_{i=1}^{N-1} (-1)^i\, Z^{N-i}\,{U}_{i}(-\kfive) + (-1)^N \Big)^2 
- 4(\beta\Lambda)^{2N} Z^{-(-\kfive)}
\end{equation}
and interpret it as the curve describing an SU($N$) theory with Chern-Simons coupling 
$-\kfive$, since it has exactly the same form of (\ref{SWcurvek5d}). 
At the classical level, {\it{i.e.}} for $\beta\Lambda\to 0$, it is trivial to check 
that $\widetilde{U}_{N-i}^{\mathrm{class}}=U_i^{\mathrm{class}}$.
Indeed, it suffices to perform the inversion of $A_u$ in (\ref{Ucl}) and use the tracelessness condition
(\ref{SUNA}). What is less obvious is to check the relation (\ref{UUtilde}) at the quantum level, {\it{i.e.}}
when the non-perturbative corrections are taken into account. In Appendix~\ref{chiralcorr} 
we explicitly verify this relation exploiting the localization calculation of the chiral 
correlators at 1-instanton. 
This provides a clear confirmation of the fact that the Seiberg-Witten curve takes the form
(\ref{SWcurvek5d}) for negative Chern-Simons levels also. As a bonus, we see that
the constraint (\ref{cond2k5d}) has a natural interpretation also from the 
point of view of the Seiberg-Witten curve. 

We now turn to the resolvent of the 5d gauge theory. This is the 
generating function of all the chiral correlators and is defined as the following expectation 
value \cite{Wijnholt:2004rg}:
\begin{equation}
{T} = \Big\langle \Tr\coth \frac{\beta(z-\Phi)}{2}\Big\rangle = \frac{2}{\beta}\frac{\p}{\p z} 
\Big\langle \Tr\log \Big(2\sinh \frac{\beta(z-\Phi)}{2}\Big) \Big\rangle
\label{resolventT}
\end{equation}
where $\Phi$ is the complex scalar field of the adjoint vector multiplet. Setting
\begin{equation}
z=\frac{1}{\beta}\log Z
\label{zZ}
\end{equation}
and expanding for large $Z$, we find
\begin{equation}
\label{TCCR}
{T} = N +2 \sum_{\ell=1}^{\infty} \frac{V_{\ell}}{Z^{\ell}}
\end{equation}
where
\begin{equation}
V_{\ell} = \Big \langle \Tr \rme^{\ell\,\beta\,\Phi}\Big\rangle ~.
\end{equation}
Of course, due to the SU($N$) condition (\ref{SUNA}), only the correlators $V_\ell$ with $\ell=1,\cdots,N-1$
are independent of each other.

We propose that the integral of the resolvent is given by
\begin{equation}
\label{integralresolvent}
\Big\langle \Tr\log \Big(2\sinh \frac{\beta(z-\Phi)}{2}\Big) \Big\rangle 
= \log\Big( \frac{P_N(Z) + Y}{2}\Big)
\end{equation}
where $Y$ satisfies the Seiberg-Witten curve in \eqref{SWcurvek5d}, and $z$ is related to $Z$ 
as in (\ref{zZ}). This proposal is suggested by the fact that the quantity appearing on the right hand
side is closely related to  the Seiberg-Witten differential of the 5d gauge 
theory \cite{Brini:2008rh}. 
Differentiating (\ref{integralresolvent}) with respect to $z$, after a straightforward calculation we 
obtain the explicit expression for the resolvent in terms of the function appearing in the Seiberg-Witten 
curve, namely
\begin{equation}
\label{5dresolventk}
\begin{aligned}
{T} &= \frac{2}{\beta}\frac{\p}{\p z} \bigg[\log\Big( \frac{P_N(Z) + Y}{2}\Big)\bigg] = 
2\,Z \,\frac{{P}_N^{\,\prime}(Z)}{Y} - \kfive\,\Big(1- \frac{{P}_N(Z)}{Y}\Big)
\end{aligned}
\end{equation}
where $^\prime$ stands for the derivative with respect to $Z$. 
The first term is precisely the 5d lift of the classic result from \cite{Cachazo:2002ry}, which 
was already used in \cite{Ashok:2017lko} for the case $\kfive=0$. 
The second term is the modification due to the Chern-Simons coupling. 

Inserting (\ref{SWcurvek5d}) and (\ref{PNZexp}) into the right hand side 
of (\ref{5dresolventk}) and expanding 
for large $Z$, we obtain an expression for the resolvent in terms of the gauge invariant 
coordinates $U_i(\kfive)$. This can then be compared with (\ref{TCCR}) to establish a
relation with the chiral correlators $V_\ell$. Proceeding this way, we find for example
\begin{equation}
U_1(\kfive) = V_1-(\beta\Lambda)^{2N}\,\delta_{\,\kfive,1-N}~.
\label{U1V1}
\end{equation}
Similar relations can be found for the higher $U_i(\kfive)$'s as we show explicitly 
in Appendix~\ref{chiralcorr}.
There, we also show that the correlators $V_\ell$ can be calculated order by order in the instanton 
expansion using localization methods involving the partition function (\ref{ZY5d}) with suitable insertions. 
Thus, our proposal for 
the resolvent provides a systematic way to obtain the explicit non-perturbative 
expressions for $U_i(\kfive)$ in terms of the Coulomb vacuum expectation values of the 
5d gauge theory. These can be used to check the relation
(\ref{UUtilde}), thus confirming the consistency of the whole construction. 
Furthermore, as we will see in the 
next section, this knowledge will prove to be an essential ingredient to study surface operators as 
coupled 3d/5d gauge theories.

\section{5d gauge theories with surface operators}
\label{localization}

We now turn to the study of SU$(N)$ gauge theories on $\mathbb{R}^4\times S^{1}$ 
in the presence of a surface operator extended along a plane $\mathbb{R}^2\subset 
\mathbb{R}^4$ and wrapped around $S^1$. We treat such surface operators as 
monodromy defects, also known as Gukov-Witten defects \cite{Gukov:2006jk, Gukov:2008sn}.
The discrete data that label these defects are the partitions of $N$, {\it{i.e.}} the sets of
positive integers $\vec{n}=[n_1, n_2, \ldots \, , n_M]$ such that $\sum_{i=1}^M n_i=N$. They are related
to the breaking pattern (or Levi decomposition) of the gauge group near the defect as follows,
\begin{equation}
\mathrm{SU}(N)~~\longrightarrow~~\mathrm{S}\big[\mathrm{U}(n_1)\times\ldots\times\mathrm{U}(n_M)\big]~.
\label{Levi}
\end{equation}
The instanton partition function in the presence of such a defect  
can be obtained by generalizing the pure five-dimensional analysis presented in the 
previous section with the addition of 
a $\mathbb{Z}_M$ orbifold projection \cite{Kanno:2011fw}, along
the lines discussed in \cite{Ashok:2017lko} in the absence of Chern-Simons interactions. 
The result is the partition function for the so-called ramified instantons.

\subsection{Ramified instantons}
\label{localizecalc}
Let us introduce a partition of order $M$ and, for each sector $I=1,\cdots,M$, consider $d_I$ ramified
instantons\,\footnote{Here and in the following, the index $I$ is always taken modulo $M$.\label{moduloM}}.
The partition function for such a configuration can be written as
\begin{equation}
Z_{\text{inst}}[\vec{n}] = \sum_{\{d_I\}}Z_{\{d_I\}}[\vec{n}]
\label{Zramif}
\end{equation}
where
\begin{equation}
Z_{\{d_I\}}[\vec{n}]= \prod_{I=1}^M \bigg[\frac{(-q_I)^{d_I}}{d_I!}\,
\int_{\mathcal{C}} \,\prod_{\sigma=1}^{d_I} 
\Big(\beta\,\frac{d\chi_{I,\sigma}}{2\pi\ii}\Big)\, 
\rme^{-\beta\,\mathsf{m}_I \chi_{I,\sigma} } \bigg]~
z_{\{d_I\}}
\label{Zso4d5d}
\end{equation}
with
\begin{align}
z_{\{d_I\}} & = \,\prod_{I=1}^M \bigg[\prod_{\sigma,\tau=1}^{d_I}\,
\frac{1}{g\big(\chi_{I,\sigma} - \chi_{I,\tau} + \epsilon_1\big)}
\prod_{\substack{\sigma,\tau=1\\ \sigma\neq\tau}}^{d_I}\,g\big(\chi_{I,\sigma} - \chi_{I,\tau}\big)\bigg]\notag\\
&~~\times\prod_{I=1}^M \prod_{\sigma=1}^{d_I}\prod_{\rho=1}^{d_{I+1}}\,
\frac{g\left(\chi_{I,\sigma} - \chi_{I+1,\rho} + \epsilon_1 + \hat\epsilon_2\right)}
{g\left(\chi_{I,\sigma} - \chi_{I+1,\rho} + \hat\epsilon_2\right)}\label{zexplicit5d}\\
&~~\times
\prod_{I=1}^M \bigg[\prod_{\sigma=1}^{d_I} \prod_{s=1}^{n_I}
\frac{1}
{g\left(a_{I,s}-\chi_{I,\sigma} + \frac 12 (\epsilon_1 + \hat\epsilon_2)\right)}
\prod_{t=1}^{n_{I+1}}
\frac{1}
{g\left(\chi_{I,\sigma} - a_{I+1,t} + \frac 12 (\epsilon_1 + \hat\epsilon_2)\right)}\bigg]~.
\notag
\end{align}
Here $q_I$ is the instanton weight in the $I$-th sector, and 
$\hat\epsilon_2=\epsilon_2/M$ as a consequence of the $\mathbb{Z}_M$ orbifold projection.
Note that these expressions are the same as those in \cite{Ashok:2017lko}, apart from a
minor modification in the integrand of (\ref{Zso4d5d}) represented by exponential factors 
that introduce a coupling to $\Tr \chi_{I}$ with coefficient $\mathsf{m}_I$.
Anticipating the description of surface operators from the point of view of 
3d/5d coupled theories \cite{Gaiotto:2009fs, Gaiotto:2013sma}, we propose (\ref{Zramif}) and 
(\ref{Zso4d5d}) to be the generalization of the results of \cite{Ashok:2017lko} 
when Chern-Simons terms are included in the 3d gauge theories defined 
on the world-volume of the defects. We will provide strong evidence for this in the following sections.

We now describe how to evaluate the integrals (\ref{Zso4d5d}) over $\chi_{I,\sigma}$.
The procedure is quite similar to what we saw in the previous section. The first step
is the choice of the integration contour and the prescription to pick up the poles
of the integrand (\ref{zexplicit5d}). A convenient way to classify the possible contours of interest 
is via the Jeffrey-Kirwan (JK) parameter $\eta$ \cite{JK1995}. Different choices of $\eta$ correspond 
to picking different sets of poles in (\ref{Zso4d5d}), which may lead
to different results for the instanton partition function.
Such issues become even more subtle once we introduce the parameters $\mathsf{m}_I$, since non-trivial 
residues at zero or infinity, and even branch cuts may appear. 
We now describe the two choices of contour that were already
introduced in \cite{Ashok:2017lko,Gorsky:2017hro}.

\subsubsection*{Prescription JK$_{\text{I}}$}

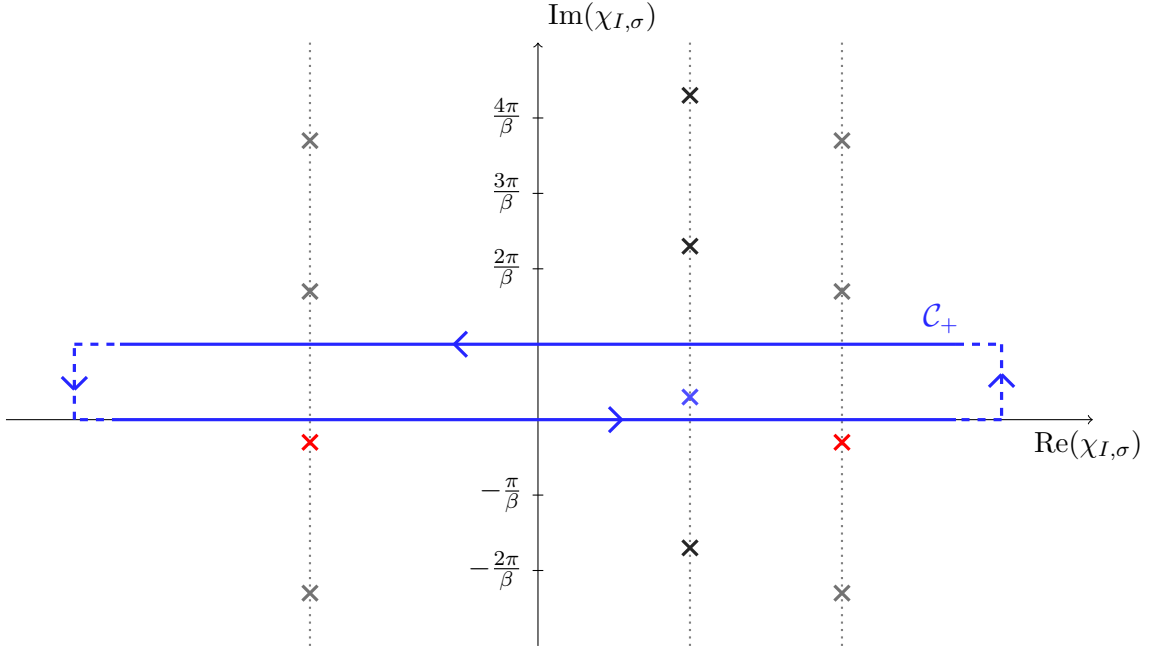
\begin{figure}[ht]
\begin{tikzpicture}[decoration={
markings,
mark=at position 0.4 with {\draw[solid] (1ex,-1ex) -- (0ex,0ex);
                \draw[solid] (1ex,1ex) -- (0ex,0ex);}}]
\draw[thin,->] (-7,0) -- (7.3,0) node[anchor=80] {$\mathrm{Re}(\chi_{I,\sigma})$ };
\draw[thin,->] (0,-3) -- (0,5) node[anchor=south west] {$\mathrm{Im}(\chi_{I,\sigma})$};
\foreach \y in {2,3,4}
    \draw (2pt,\y cm) -- (-2pt,\y cm) node[anchor=east] {$\frac{\y \pi}{\beta}$};
\draw (2pt,-1) -- (-2pt,-1) node[anchor=east] {$-\frac{\pi}{\beta}$};  
\draw (2pt,-2) -- (-2pt,-2) node[anchor=east] {$-\frac{2\pi}{\beta}$};  

\foreach \x in {2,4,-3}
\draw[gray,thick,dotted] (\x,-3)--(\x,5); 

\draw[very thick,blue!85] (-5.5,1) --(5.5,1)[postaction={decorate}] node[anchor=-60] {{\large $\mathcal{C}_+$} };
\draw[very thick,blue!85,dashed] (5.5,1) --(6.1,1); 
\draw[very thick,blue!85,dashed] (6.1,1) --(6.1,0)[postaction={decorate}]; 
\draw[very thick,blue!85,dashed] (6.1,0) --(5.5,0); 
\draw[very thick,blue!85] (5.5,0) --(-5.5,0)[postaction={decorate}];  
\draw[very thick,blue!85,dashed] (-5.5,0) --(-6.1,0); 
\draw[very thick,blue!85,dashed] (-6.1,0) --(-6.1,1)[postaction={decorate}]; 
\draw[very thick,blue!85,dashed] (-6.1,1) --(-5.5,1); 

\foreach \i in {-2,2,4}
   \draw ($(2,0.3+\i)$) node[cross,black!85,very thick]{}
   ($(4,-.3+\i)$) node[cross,black!55, very thick]{} 
   ($(-3,-.3+\i)$) node[cross,black!55,very thick]{};
\draw ($(2,0.3)$) node[cross,blue!70, very thick]{}   
($(4,-.3)$) node[cross,red,very thick]{} 
   ($(-3,-.3)$) node[cross,red,very thick]{};
\end{tikzpicture}
\caption{The contour ${\mathcal C}_+$ for the case of the $[1,2]$ surface operator in SU(3) at 1-instanton.}
\label{FigureCpluswithMs}
\end{figure}

In our first prescription for the integration contour, the JK parameter is\,\footnote{For details see for example
\cite{Gorsky:2017hro,Ashok:2017lko}.}
\begin{align}
\label{JKoriginal}
\eta=-\sum_{I=1}^{M-1}\chi_{I}+\xi\,\chi_{M}
\end{align}
with $\xi$ an arbitrary large positive number.
Using (\ref{e1e2ImaginaryPart}), one can see that this choice is equivalent to selecting the poles for 
$\chi_{I,\sigma}$ as follows
\begin{equation}
\begin{aligned} 
\label{JK1Poles}
0 &<\text{Im}\, \chi_{I,\sigma} < \frac{\pi}{\beta}\qquad\text{for}\quad I = 1, \ldots, 
M-1\quad\text{and}\quad \sigma=1,\ldots, d_I~,\\
-\frac{\pi}{\beta} &<\text{Im}\, \chi_{M,\sigma} < 0 \qquad\text{for}\quad \sigma=1,\ldots, d_M~.
\end{aligned}
\end{equation}
This corresponds to choosing the contour ${\mathcal C}_+$ for the first $M-1$ sets of integration 
variables and the contour ${\mathcal C}_-$ for the $M$th set. The two contours $\mathcal{C}_+$
and $\mathcal{C}_-$ are shown, respectively, in Fig.~\ref{FigureCpluswithMs} and 
Fig~\ref{FigureCminuswithMs}, for the SU(3) theory in the presence of the [1,2] surface operator, 
at the 1-instanton level.
\begin{figure}[ht]
\begin{tikzpicture}[decoration={
markings,
mark=at position 0.4 with {\draw[solid] (1ex,-1ex) -- (0ex,0ex);
                \draw[solid] (1ex,1ex) -- (0ex,0ex);}}]
\draw[thin,->] (-7,0) -- (7.3,0) node[anchor=80] {$\mathrm{Re}(\chi_{I,\sigma})$ };
\draw[thin,->] (0,-3) -- (0,5) node[anchor=south west] {$\mathrm{Im}(\chi_{I,\sigma})$};
\foreach \y in {2,3,4}
    \draw (2pt,\y cm) -- (-2pt,\y cm) node[anchor=east] {$\frac{\y \pi}{\beta}$};
    \draw (2pt,-2 cm) -- (-2pt,-2 cm) node[anchor=east] {$-\frac{2 \pi}{\beta}$};
\draw (2pt,1) -- (-2pt,1) node[anchor=east] {$\frac{\pi}{\beta}$};  

\foreach \x in {2,4,-3}
\draw[gray,thick,dotted] (\x,-3)--(\x,5); 

\draw[very thick,red] (-5.5,-1) --(5.5,-1)[postaction={decorate}] ;
\draw[very thick,red,dashed] (5.5,-1) --(6.1,-1); 
\draw[very thick,red,dashed] (6.1,-1) --(6.1,0)[postaction={decorate}]; 
\draw[very thick,red,dashed] (6.1,0) --(5.5,0); 
\draw[very thick,red] (5.5,0) node[anchor=-60] {{\large $\mathcal{C}_-$} }--(-5.5,0)[postaction={decorate}];  
\draw[very thick,red,dashed] (-5.5,0) --(-6.1,0); 
\draw[very thick,red,dashed] (-6.1,0) --(-6.1,-1)[postaction={decorate}]; 
\draw[very thick,red,dashed] (-6.1,-1) --(-5.5,-1);
\foreach \i in {-2,2,4}
   \draw ($(2,0.3+\i)$) node[cross,black!85,very thick]{}
   ($(4,-.3+\i)$) node[cross,black!55, very thick]{} 
   ($(-3,-.3+\i)$) node[cross,black!55,very thick]{};
\draw ($(2,0.3)$) node[cross,blue!70, very thick]{}   
($(4,-.3)$) node[cross,red,very thick]{} 
   ($(-3,-.3)$) node[cross,red,very thick]{};
\end{tikzpicture}
\caption{The contour ${\mathcal C}_-$ for the case of the $[1,2]$ surface operator in SU(3) at 1-instanton}
\label{FigureCminuswithMs}
\end{figure}

\subsubsection*{Prescription JK$_{\text{II}}$}
In our second prescription  the JK parameter is given by
\begin{equation}
\label{JKDual}
\widetilde{\eta}=\sum_{I=1}^{M-1}\chi_I-\xi\,\chi_M 
 \end{equation}
 where again $\xi$ is an arbitrary large positive number.
 This corresponds to choosing the poles as follows
 \begin{equation}
\begin{aligned} 
\label{JK2Poles}
-\frac{\pi}{\beta} &<\text{Im}\, \chi_{I,\sigma} < 0~~
\qquad\text{for}\quad I = 1, \ldots, M-1\quad\text{and}\quad \sigma=1,\ldots, d_I~,\\
0 &<\text{Im}\, \chi_{M,\sigma} < \frac{\pi}{\beta}\qquad\text{for}\quad \sigma=1,\ldots, d_M~.
\end{aligned}
\end{equation}
Equivalently we can say that one selects the contour ${\mathcal C}_-$ (see Fig.~\ref{FigureCminuswithMs}) 
for the first $M-1$ sets of $\chi$-variables and the contour ${\mathcal C}_+$ 
(see Fig.~\ref{FigureCpluswithMs}) for the $M$th set. 
This prescription is clearly complementary to the first one. 

Our goal is to understand how and when the two prescriptions JK$_{\text{I}}$ and JK$_{\text{II}}$ can be
related by contour deformation so that the partition functions obtained via the two match.
To illustrate this point it suffices again to focus on the 1-instanton case.

\subsection{The 1-instanton partition function}
\label{subsecn:1inst}

Let us consider the 1-instanton contribution to the partition function for a general surface operator 
of type $\vec{n}=[n_1, n_2, \ldots \, , n_M]$. To express the formulas 
in a compact form, it is convenient to introduce 
the integers
\begin{equation}
r_I = \sum_{J=1}^I n_J
\label{rI}
\end{equation}
which will be used also in Section~\ref{secn:3d5d}. We also choose an ordering such that the 
Coulomb vacuum expectation values are partitioned as follows
\begin{equation}
\label{asplit}
\big\{a_1,\ldots,  a_{r_1}|\ldots \big|
a_{r_{I-1}+1}, \ldots a_{r_{I}}\big|
a_{r_{I} + 1}, \ldots a_{r_{I+1}}\big|
\ldots |a_{r_{M-1} +1}, \ldots, a_N \big\} ~.  
\end{equation}
{From} the definition (\ref{rI}), it is clear that each partition is of length $n_I$. 
Compared to the notation we have used in (\ref{Zso4d5d}), this ordering corresponds to
\begin{equation}
a_{I,s} = a_{r_{I-1} + s}\quad\text{for}\quad s = 1, \ldots n_I
\end{equation}
with the understanding that $r_0=0$. Using this notation, the 1-instanton partition function in the presence
of a generic surface operator becomes
\begin{equation}
\begin{aligned}
\label{Partition_function_gen_op}
Z_{1-\text{inst}} &= 
               -\frac{1}{g(\epsilon_1)}\sum_{I=1}^{M} q_I\int_\mathcal{C} 
\Big(\beta\,\frac{d\chi_I}{2\pi\ii}\Big) ~\rme^{- \beta \mathsf{m}_I \chi_I}\!\!
\prod_{\ell= r_{I-1}+1}^{r_I}\,\frac{1}{g\big(a_{\ell}-\chi_I+\frac{1}{2}(\epsilon_1+\hat{\epsilon}_2)\big)} \\
& \hspace{2.5cm} \times \prod_{j= r_{I}+1}^{r_{I+1}} 
\frac{1}{g\big(\chi_I-a_{j} +\frac{1}{2}(\epsilon_1+\hat{\epsilon}_2) \big)}~. 
\end{aligned}
\end{equation}
We now perform a change of variables as in (\ref{bigvars}) to obtain $X_I$, $E_1$ and $\hat E_2$ 
from $\chi_I$, $\epsilon_1$ and $\hat\epsilon_2$ respectively.
In these new variables, after some simple manipulations,
(\ref{Partition_function_gen_op}) can be brought into the following form
\begin{equation}
\begin{aligned}
\label{Partition_function_gen_op2}
Z_{1-\text{inst}}&=- \sum_{I=1}^{M} {q_I}\,
\frac{E_1^{\frac {n_{I} + n_{I+1}}4+\frac{1}{2}}\, \hat{E}_2^{\frac {n_{I} + n_{I+1}}4}}{E_1 - 1}
\int_{\mathcal{C}} \frac{dX_I}{2\pi\ii} ~X_I^{\frac {n_{I} + n_{I+1}}2 - \mathsf{m}_I-1}\\
&\hspace{1.2cm}\times\prod_{\ell =r_{I-1}+1}^{r_I}
\frac{1}{\Big(A_{\ell}\, \sqrt{E_1\hat E_2\phantom{|}}-X_I\Big)}
 \times \prod_{j=r_{I}+1}^{r_{I+1}}\frac{1}{\Big(A_j -  X_I\,\sqrt{E_1\hat E_2\phantom{|}}\Big)} ~.
\end{aligned}
\end{equation}
{From} this explicit expression it is clear that, besides the simple poles at
\begin{equation}
X_I=\,A_\ell\,\sqrt{E_1\hat E_2\phantom{}}\quad\quad{\mbox{and}}\quad \quad
X_I=\,\frac{A_j}{\sqrt{E_1\hat E_2\phantom{|}}}~,
\label{polesXI}
\end{equation}
the integrand may possess branch cuts as well as singularities at $X_I=0$ and $X_I=\infty$ 
depending on the value of $m_I$. If this is the case,
the two contours are obviously not equivalent to each other. To avoid branch cuts we must require that
\begin{equation}
\label{cond1mI}
 \mathsf{m}_I + \frac{n_{I} + n_{I+1}}{2}~\in\, \mathbb{Z}\quad\text{for}\quad I =1, \ldots, M  
\end{equation}
where $n_{M+1} = n_1$ (see footnote~\ref{moduloM}). Furthermore, to avoid contributions from the 
nonphysical singularities at $X_I=0$ and $X_I=\infty$, we must constrain $\mathsf{m}_I$ such that
\begin{equation}
\label{cond2mI}
 \left\vert \mathsf{m}_I \right\vert \leq \,\frac {n_I + n_{I+1}}2 -1\quad\text{for}\quad I =1, \ldots, M ~.
\end{equation}
When conditions (\ref{cond1mI}) and (\ref{cond2mI}) are satisfied, the two JK prescriptions 
lead to the same result because the contours $\mathcal{C}_+$ and $\mathcal{C}_-$ can be smoothly 
deformed into each other.

The above analysis can be repeated at higher instanton levels, but the explicit expressions quickly 
become rather involved and not very illuminating. We have performed several explicit calculations up to
three instantons in theories with low rank gauge groups and have encountered no other constraints 
on $\mathsf{m}_I$ other than those in (\ref{cond1mI}) and (\ref{cond2mI}) 
in order to obtain results that are independent of the prescription used to evaluate the integrals.

\subsection{Parameter map}
Once the instanton partition function is computed, one can extract from it the 
non-perturbative prepotential
${\mathcal F}_{\text{inst}}$ and the twisted superpotential ${\mathcal W}_{\text{inst}}$ 
according to \cite{Alday:2009fs, Alday:2010vg}
\begin{equation}
\label{ExtractW}
\log Z_{\text{inst}}=-\frac{{\mathcal F}_{\text{inst}}}{\epsilon_1\hat{\epsilon}_2}+
\frac{{\mathcal W}_{\text{inst}}}{\epsilon_1}+\text{regular terms}~.
\end{equation}
The prepotential governs the dynamics of the bulk 5d theory and depends on the 
parameters of this theory, namely the vacuum expectation values of the adjoint scalars, 
the Chern-Simons coupling $\kfive$ and the instanton
counting parameter $q$.
The twisted superpotential, instead, controls the dynamics on the surface operator and in addition to these depends
on the parameters that label the defect. From our explicit results we have verified
that ${\mathcal F}_{\text{inst}}$ depends only on the vacuum expectation values, the 
sum of all $\mathsf{m}_I$, and the product of all $q_I$. In particular these latter 
combinations play the role, respectively, of $\kfive$ and $q$; thus, comparing
with what we have seen in Section~\ref{secn:5dCS}, we are led to
\begin{align}
\kfive &=\, \sum_{I=1}^M\mathsf{m}_I
\label{k5dissum}~,\\
q&=\,\prod_{I=1}^M q_I~=\,(-1)^N(\beta \Lambda)^{2N}~.
\label{qisproduct}
\end{align}

We recall that the instanton counting parameters $q_I$ are related to the monodromy properties 
of the SU($N$) gauge connection once the breaking pattern (\ref{Levi}) is enforced by 
the presence of the defect. 
Building on earlier works \cite{Alday:2010vg,Kanno:2011fw}, this fact was explicitly shown 
in \cite{Ashok:2017odt} for the $\mathcal{N}=2^\star$ theories, and already 
used in \cite{Ashok:2017lko} for the pure $\mathcal{N}=2$ theories
(see for instance, Eq.~$(2.48)$ in \cite{Ashok:2017lko}). 
Notice that only the product of all $q_I$'s has a global 5d interpretation as is clear from (\ref{qisproduct}).
Similarly, the parameters $\mathsf{m}_I$ describe how the Chern-Simons level $\kfive$ of the 5d SU($N$) theory is split among the $M$ factors in the Levi decomposition (\ref{Levi}) and as such they are part of the data that specify the defect. {From} (\ref{zexplicit5d}) we see that these parameters 
appear like Chern-Simons couplings for the U($n_I$) factors, even though one should take 
into account that the ramified instanton partition function (\ref{Zso4d5d}) is not factorizable 
into a product of $M$ partition functions.
Finally, we observe that the constraints (\ref{cond1mI}) 
and (\ref{cond2mI}) imply that 
\begin{equation}
\kfive \in \mathbb{Z}\quad\quad\mbox{and}\quad\quad
\left\vert\,\kfive\right\vert\leq N-M~.
\label{k5dconstraintSO}
\end{equation}

\subsection{Simple surface operators}
\label{subsecn:simple}

For the purpose of illustration, we now consider in detail the case of the simple surface operator 
of type $[1,N-1]$ in the SU($N$) theory. This case corresponds to setting $M=2$ and splitting
the classical vacuum expectation values as $\big\{a_1\big| a_2, \ldots, a_N\big\}$.
Using this in (\ref{Partition_function_gen_op}), the 1-instanton contribution to the partition function
becomes
\begin{equation}
\begin{aligned}
\label{Zk1simple}
Z_{1-\text{inst}} =& -\frac{q_1}{g(\epsilon_1)}\int_{\mathcal{C}}\! \Big(\beta \,\frac{d\chi_1}{2\pi\ii}\Big) 
\frac{\rme^{-\beta \mathsf{m}_1 \chi_1}}{g\big(a_1-\chi_1+\frac{1}{2}(\epsilon_1+\hat{\epsilon}_2)\big)}
~ \prod_{i=2}^N \frac{1}{g\big(\chi_1-a_i +\frac{1}{2}(\epsilon_1+\hat{\epsilon}_2) \big)}\\
&-\frac{q_2}{g(\epsilon_1)}\int_{\mathcal{C}}\!  \Big(\beta \,\frac{d\chi_2}{2\pi\ii}\Big)
\frac{\rme^{-\beta \mathsf{m}_2 \chi_2}}{g\big(\chi_2-a_1+\frac{1}{2}(\epsilon_1+\hat{\epsilon}_2)\big)}
~ \prod_{i=2}^N \frac{1}{g\big(a_i-\chi_2 +\frac{1}{2}(\epsilon_1+\hat{\epsilon}_2)\big)}
\end{aligned}
\end{equation}
We now evaluate the integrals over $\chi_1$ and $\chi_2$ using the two JK prescriptions described above.
In the first prescription JK$_{\text{I}}$, according to (\ref{JK1Poles}), the contributing poles are 
located at
\begin{equation}
\chi_1 = a_1 +\frac{1}{2}(\epsilon_1+\hat{\epsilon}_2)\qquad\text{and}\qquad \chi_2 = a_1 
- \frac{1}{2}(\epsilon_1+\hat{\epsilon}_2)~.
\end{equation}
Calculating the corresponding residues, extracting the twisted 
superpotential by means of (\ref{ExtractW}), and expressing the results in terms of the 
variables (\ref{bigvars}), we find
\begin{equation}
\label{W1instsimpleJK1}
{\mathcal W}^{\,(\text{I})}_{1-\text{inst}} =  \frac{1}{\beta}\,
\Big(q_1\, A_1^{\frac{N}{2}-\mathsf{m}_1-1}+(-1)^{N-1}q_2\,A_1^{\frac{N}{2}-\mathsf{m}_{2}-1}
\Big)\prod_{i=2}^N(A_1-A_i)^{-1}~.
\end{equation}
Next we consider the second prescription JK$_{\text{II}}$; according to (\ref{JK2Poles}), the contributing
poles are located at
 \begin{equation}
 \chi_1 = a_u - \frac{1}{2}(\epsilon_1+\epsilon_2)\qquad\mbox{and}\qquad 
 \chi_2 = a_u + \frac{1}{2}(\epsilon_1+\epsilon_2)\qquad\text{for}~~ u=2,3, \ldots, ~.
 \end{equation}
The corresponding twisted superpotential is
\begin{equation}
\label{W1instsimpleJK2}
{\mathcal W}^{\,(\text{II})}_{1-\text{inst}}
= -  \frac{1}{\beta}\sum_{i=2}^N \bigg[\Big(
q_1\, A_i^{\frac{N}{2}-\mathsf{m}_1-1}+
(-1)^{N-1}q_2\,A_i^{\frac{N}{2}-\mathsf{m}_{2}-1}\Big) \!\!
\prod_{j=1\,,\,j\ne i}^N(A_i-A_j)^{-1}\bigg]~. 
\end{equation}

Comparing the two expressions (\ref{W1instsimpleJK1}) and (\ref{W1instsimpleJK2}), we see that they are 
very different from each other. However, if we use the SU($N$) condition (\ref{SUNA}) and 
impose the constraints (\ref{cond1mI}) and (\ref{cond2mI}), which for $M=2$ are
\begin{equation}
\mathsf{m}_{1,2}+\frac{N}{2} \in \mathbb{Z}\quad\quad\mbox{and}\quad\quad
\vert\,\mathsf{m}_{1,2}\,\vert\le \frac{N}{2}-1~,
\label{m12}
\end{equation}
one can verify that 
${\mathcal W}^{\,(\text{I})}_{1-\text{inst}}$ and ${\mathcal W}^{\,(\text{II})}_{1-\text{inst}}$ 
match. We have explicitly checked that the
match continues to hold at higher instanton levels 
(up to three instantons for the low rank SU$(N)$ theories).

\section{3d/5d quiver theories with Chern-Simons terms}
\label{secn:3d5d}

We now study surface operators from the point of view of 3d/5d coupled systems compactified 
on a circle of radius $\beta$, by extending the analysis of \cite{Ashok:2017lko} to explicitly 
include Chern-Simons interactions \footnote{The brane construction of 3d gauge theories with Chern Simons interactions has been studied in \cite{Kitao:1998mf}, \cite{Bergman:1999na}.}. We then derive and solve the resulting twisted chiral ring equations.

\subsection{The linear quiver and its twisted chiral ring equations}
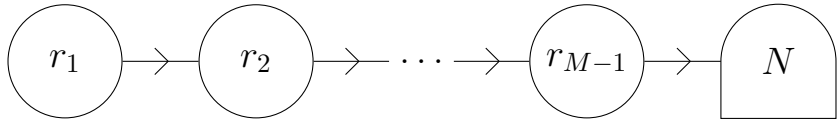
\begin{figure}[ht]
\begin{center}
\begin{tikzpicture}[decoration={markings, mark=at position 0.6 with {\draw (-5pt,-5pt) -- (0pt,0pt);       
         \draw (-5pt,5pt) -- (0pt,0pt);}}]
  \matrix[row sep=10mm,column sep=5mm] {
      \node (g1)[gauge] {\Large $r_1$};  & & \node(g2)[gauge] {\Large $r_2$};  && \node(dots){\Large 
      $\ldots$};
      & & \node(glast)[gauge] {\Large $r_{M-1}$};& &\node(gfN)[gaugedflavor]{\Large $N$};\\
  };
\graph{(g1) --[postaction={decorate}](g2) --[postaction={decorate}](dots)--[postaction={decorate}](glast) 
--[postaction={decorate}](gfN);};
\end{tikzpicture}
\end{center}
\vspace{-0.5cm}
\caption{The quiver that describes the generic surface operator in pure SU$(N)$ gauge theory.}
\label{quiverpicgeneric1}
\end{figure}

Our proposal is that the 3d/5d system that corresponds to a surface operator 
labeled by the partition $\vec{n}=[n_1, n_2, \ldots \, , n_M]$ and treated with the first JK prescription
(\ref{JKoriginal}), is the quiver theory described in Fig.~\ref{quiverpicgeneric1}. 
Here, the circular nodes represent 3d U$(r_I)$ gauge theories, the rightmost node represents a 
5d SU$(N)$ gauge theory, and the arrows denote bifundamental matter fields. 
The ranks $r_I$ of the 3d gauge groups are related to the surface operator data $n_I$ as in (\ref{rI}). 

The gauge degrees of freedom in each node can be organized in an adjoint twisted chiral multiplet 
$\Sigma^{(I)}$, which for notational simplicity we will often denote by its lowest scalar 
component $\sigma^{(I)}$. 
The low-energy effective action on the Coulomb moduli space is parameterized by the diagonal 
components of $\sigma^{(I)}$:
\begin{equation}
\sigma^{(I)}=\text{diag}\big\{\,\sigma^{(I)}_1,\sigma^{(I)}_2,\ldots,\sigma^{(I)}_{r_I}\,\big\}~.
\end{equation}
This can be obtained by integrating out the matter multiplets 
corresponding to the arrows of the quiver, which are generically 
massive when $\sigma^{(I)}$ and the adjoint scalar field $\Phi$ of the SU$(N)$ theory 
acquire non-vanishing vacuum expectation values \cite{Hanany:1997vm}. 
Supersymmetry implies that the effective action can be encoded in a twisted chiral superpotential, 
which takes the form (see \cite{Ashok:2017lko} for details):
\begin{equation}
\begin{aligned}
{\mathcal W}_0=\sum_{I=1}^{M-1}\sum_{s=1}^{r_I}&b_I \log (\beta\Lambda_I\!)\,
\sigma^{(I)}_s
- \!\sum_{I=1}^{M-2} \sum_{s=1}^{r_{I}} \sum_{t=1}^{r_{I+1}}
\ell\big(\sigma^{(I)}_s- \sigma^{(I+1)}_t\big)
-\!\!\sum_{s=1}^{r_{M-1}}\!\Big\langle 
\Tr \ell\big(\sigma^{(M-1)}_s - \Phi\big) \Big\rangle
\end{aligned}
\label{tildeW15d}
\end{equation}
where $\Lambda_I$ is the (complexified) IR scale of the $I$-th node and
\begin{equation}
\label{defbI}
b_I=r_{I+1}-r_{I-1}
\end{equation}
for $I=1,\cdots,M-1$\,\footnote{Here and in the following we understand
that $r_0=0$ and $r_M=N$.}. 
The expectation value in the last term of (\ref{tildeW15d}) is taken in the 5d SU($N$) 
gauge theory and the function $\ell(z)$ obeys the relation
\begin{equation}
\label{derell}
\partial_z\ell(z)=\log\Big(2 \sinh \frac{\beta z}{2}\Big)\,.
\end{equation}
In each 3d node of the quiver we can turn on a Chern-Simons term with coupling $\mathsf{k}_{I}$. 
Upon circle compactification, these Chern-Simons terms give rise to an additional term in the superpotential
which is \cite{Chen:2012we}\,\footnote{This differs from the conventions in our previous paper 
\cite{Ashok:2017lko} by a sign.}
\begin{equation}
{\mathcal W}^{(I)}_{\mathrm{CS}}= -\frac{\beta\,\mathsf{k}_I }{2}\,\Tr \big(\sigma^{(I)}\big)^2 ~.
\end{equation}
Thus the complete twisted superpotential governing the 3d/5d quiver theory of Fig.~\ref{quiverpicgeneric1}
is
\begin{equation}
{\mathcal W} = {\mathcal W}_0 + \sum_{I=1}^{M-1} {\mathcal W}^{(I)}_{\mathrm{CS}}~.
\end{equation}

The vacuum expectation values of the 5d fields appear in this twisted superpotential $\mathcal{W}$ 
in such a way that extremizing the latter leads to a discrete set of massive vacua, thus completely lifting 
the 3d Coulomb branch.  We now derive the so-called twisted chiral ring equations which
identify these massive vacua and study specific solutions with the aim of checking 
our proposal.
We will show that the twisted chiral superpotential evaluated in these (massive) vacua coincides 
with the one obtained using the first JK prescription in the localization analysis.

The extremization equations of the superpotential $\mathcal{W}$ 
take the following form \cite{Nekrasov:2009ui, Nekrasov:2009rc}:
\begin{equation}
\label{TCRdefn}
\exp\left(\frac{\partial {\mathcal W}}{\partial\sigma^{(I)}_s}\right)=1~.
\end{equation}
These equations were analyzed in great detail in \cite{Ashok:2017lko}, and we will be brief 
in reviewing their derivation. 
We first introduce the functions
\begin{equation}
Q_I(z)=\prod_{s=1}^{r_I}\Big(2\sinh\frac{\beta(z-\sigma^{(I)}_s)}{2}\Big)~,
 \label{QIz}
\end{equation}
or, equivalently,
\begin{equation}
Q_I(Z)= Z^{-\frac{r_I}{2}}
 \prod_{s=1}^{r_I}\big(S_s^{(I)}\big)^{-\frac{1}{2}}\,\big(Z-S_s^{(I)}\big)
 \label{QIZ}
\end{equation}
where
\begin{equation}
\label{5dexpvars}
\sigma^{(I)}_s = \frac{1}{\beta} \log S^{(I)}_s \quad \text{and}\quad
z=\frac{1}{\beta}\log Z\, .
\end{equation}
Then, for $I=1,\ldots,M-2$, the twisted chiral ring equations (\ref{TCRdefn}) become
\begin{equation}
{Q}_{{I+1}}(Z)= (-1)^{r_{I-1}}\,(\beta\Lambda_{I})^{b_I}Z\, ^{-\mathsf{k}_I}\,
{Q}_{{I-1}}(Z)
\label{CR3dv0}
\end{equation}
for $Z=S^{(I)}_s$. Here we understand that $Q_0=1$. 
For the last 3d gauge node in the quiver, {\it{i.e.}} for $I=M-1$, we obtain
\begin{equation}
\exp{\Big\langle \Tr\log\Big(2\sinh\frac{\beta(z -\Phi)}{2}\Big)\Big\rangle}
 = (-1)^{r_{M-2}}\,(\beta\Lambda_{M-1})^{b_{M-1}}\,Z^{-\mathsf{k}_{M-1}}\,{Q}_{{M-2}}(Z)\,,
\label{CR5d}
\end{equation}
for $z=\sigma^{(M-1)}_s$ or, equivalently, $Z=S^{(M-1)}_s$.
This equation clearly shows that the coupling between the 3d and the 5d theories occurs via the 
integral of the resolvent of the SU$(N)$ gauge theory 
(see (\ref{resolventT})). Using (\ref{integralresolvent}), after 
some simple algebraic manipulations, we can rewrite (\ref{CR5d}) as  
\begin{equation}
\begin{aligned}
{P}_N(Z) &=(-1)^{r_{M-2}}\,(\beta\Lambda_{M-1})^{b_{M-1}}\, 
Z^{-\mathsf{k}_{M-1}}{Q}_{{M-2}}(Z) \phantom{\bigg|}\\
&\qquad\qquad\qquad\qquad
+(-1)^{r_{M-2}} \,\frac{(\beta\Lambda)^{2N}Z^{-\kfive+\mathsf{k}_{M-1} } }
{(\beta\Lambda_{M-1})^{b_{M-1}}\, Q_{M-2}(Z) }
\end{aligned}
\label{CR5dDualfinv1}
\end{equation}
where $P_N$ is defined in (\ref{PNZexp}).

We now follow the same method described in \cite{Ashok:2017lko} and recursively solve the
chiral ring equations (\ref{CR3dv0}) and (\ref{CR5dDualfinv1}) in a semi-classical expansion 
around 
\begin{equation}
\label{Scl}
S_{\star,\mathrm{class}}^{(I)} = \text{diag}(A_1, \ldots\,, A_{r_I})~,
\end{equation}
using the perturbative ansatz
\begin{align}
\label{ansS}
S^{(I)}_\star =S_{\star,\mathrm{class}}^{(I)} +
\delta S^{(I)}_\star = \mathrm{diag}\Big(A_1+\sum_\ell \delta S^{(I)}_{1,\ell},\cdots,
A_{r_I}+\sum_\ell \delta S^{(I)}_{r_I,\ell}\Big)
\end{align}
where the increasing values of the index $\ell$ in (\ref{ansS}) correspond to corrections of increasing order 
in the compactification radius $\beta$. 
Inserting this ansatz into (\ref{CR3dv0}) and (\ref{CR5dDualfinv1}),
we can explicitly work out the solution $S^{(I)}_\star$ to the desired perturbative order 
in $\beta$, and show that the twisted superpotential evaluated on this solution, which we denote by
$\cW_\star$, can be matched with the twisted superpotential for the corresponding surface defect 
obtained via localization using the JK$_I$ prescription. For this purpose, it is more convenient to 
consider the logarithmic derivatives of $\cW_\star$ with respect to $\Lambda_I$
which have a simple expression, namely
\begin{equation}
\label{logLIder}
\Lambda_I\frac{d\cW_\star}{d\Lambda_I}
= \frac{b_I}{\beta}\,\tr \log S^{(I)}_{\star} ~.
\end{equation}
As we will see, in order to obtain agreement 
we need a precise map between the IR parameters $\Lambda_I$ and $\Lambda$ of the 3d/5d coupled 
system and the instanton counting parameters $q_I$ and $q_M$, and also a specific identification 
between the Chern-Simons levels of the 3d and 5d nodes with the parameters $\mathsf{m}_{I}$ introduced
in the localization integrand.

We now give some details, starting from the case of simple operators, which were already analyzed in
Section~\ref{subsecn:simple} from the localization point of view.

\subsubsection*{Simple surface operators}
\label{1N1usual}
\begin{figure}[ht]
\begin{center}
\begin{tikzpicture}[decoration={
markings,mark=at position 0.6 with {\draw (-5pt,-5pt) -- (0pt,0pt);
                \draw (-5pt,5pt) -- (0pt,0pt);}}]
  \matrix[row sep=10mm,column sep=5mm] {
      \node(g1)[gauge] {\Large $1$ };      & &\node(gf2)[gaugedflavor]{\Large $N$};\\
  };
\graph{(g1) --[postaction={decorate}](gf2);};
\end{tikzpicture}
\end{center}
\vspace{-0.5cm}
\caption{The quiver diagram for the simple surface operator of type $[1,N-1]$ in the SU($N$) theory.}
\label{Simplequiver}
\end{figure}
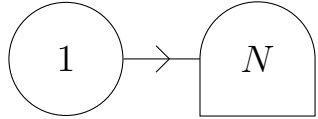

In this case there is only one 3d gauge node, and the quiver diagram is represented in
Fig.~\ref{Simplequiver}.

Correspondingly, we have just one variable $\sigma^{(1)}$, or $S^{(1)}$, 
and one chiral ring equation which is
\begin{equation}
\label{PNeqnv1}
{P}_N(S^{(1)}) = (\beta \Lambda_1)^N \big(S^{(1)}\big)^{-\mathsf{k}_1} + 
\frac{(\beta \Lambda)^{2N}}{(\beta \Lambda_1)^N} \big(S^{(1)}\big)^{-\kfive + \mathsf{k}_{1}}~.
\end{equation}
This follows from (\ref{CR5dDualfinv1}) with $M=2$, which implies $b_1=N$. 
The first non-trivial order is easy to extract. Indeed, we can start from the ansatz (\ref{Scl}), 
namely from $S^{(1)}_{\star,\mathrm{class}}= A_1$, and use the classical approximation of 
${P}_N$ given in (\ref{PNclass}) to write
\begin{equation}
{P}_N(S^{(1)}) = \big(S^{(1)}\big)^{-\frac{N}{2}}
\prod_{u=1}^N \big(S^{(1)}- A_u\big) + O\big((\beta\Lambda)^{2N}\big)~.
\end{equation}
Inserting this in (\ref{PNeqnv1}), we find
\begin{equation}
S^{(1)}_\star  = A_1\bigg[1+\Big(
(\beta\Lambda_1)^N\, A_1^{\frac{N}{2}-\mathsf{k}_1-1} + 
\frac{(\beta\Lambda)^{2N}}{(\beta\Lambda_1)^N}\,  A_1^{\frac{N}{2}-\kfive + \mathsf{k}_{1}-1}\Big)\prod_{i=2}^N 
\frac{1}{(A_1 - A_i)} + \cdots\bigg]
\end{equation}
where the ellipses stand for terms of order $(\beta\Lambda)^{4N}$ and higher.
Finally, from (\ref{logLIder}) we obtain 
\begin{align}
\frac{1}{N}\,&\Lambda_1\frac{d{\mathcal W}_{\star}}{d\Lambda_1}
=\frac{1}{\beta}\,\log S^{(1)}_\star\phantom{\Big|}\label{resLdL1}\\
&= \frac{1}{\beta}\,\log A_1\!+\! \frac{1}{\beta}\,\Big(
(\beta\Lambda_1)^N\, A_1^{\frac{N}{2}-\mathsf{k}_1-1} + \frac{(\beta\Lambda)^{2N}}{(\beta\Lambda_1)^N}\,  
A_1^{\frac{N}{2}-\kfive + \mathsf{k}_{1}-1}\Big)\prod_{i=2}^N 
\frac{1}{(A_1 - A_i)}+ \ldots ~.\nonumber
\end{align}
The non-perturbative part of this expression can be related to the superpotential 
${\mathcal W}^{\,(\text{I})}_{1-\text{inst}}$  
obtained via localization with the JK$_\text{I}$ prescription and given in (\ref{W1instsimpleJK1}). 
Indeed, upon making the following identifications
\begin{equation}
\label{q1q2Lambda}
q_1 = (\beta\Lambda_1)^N~,\qquad
q_2=(-1)^N\frac{(\beta\Lambda)^{2N}}{(\beta\Lambda_1)^N}~,
\end{equation}
and
\begin{equation}
\mathsf{m}_1 = \mathsf{k}_1~,\qquad
\mathsf{m}_{2}=\kfive -\mathsf{k}_1~,
\label{m1m2k}
\end{equation}
we find
\begin{equation}
\label{logderqWs}
\frac{1}{N}\,\Lambda_1\frac{d{\mathcal W}_{\star}}{d\Lambda_1}  = \frac{1}{\beta}\,\log A_1 
+  q_1 \frac{d{\mathcal W}^{\,(\text{I})}_{1-\text{inst}}}{dq_1}+ \cdots~,
\end{equation}
where on the right hand side the derivative with respect to $q_1$ is taken by keeping the product 
$q_1 q_2$ fixed, {\it{i.e.}} at a fixed 5d scale $\Lambda$. We remark that the identifications
(\ref{q1q2Lambda}) and (\ref{m1m2k}) imply
\begin{equation}
q_1 q_2 = (-1)^N(\beta\Lambda)^{2N}
\end{equation}
and
\begin{equation}
\kfive=\mathsf{m}_{1}+\mathsf{m}_{2}~,
\end{equation}
in perfect agreement, respectively, with (\ref{qisproduct}) and (\ref{k5dissum}) for $M=2$.

A similar analysis can be carried out at higher instanton levels. 
For instance, the two-instanton correction to (\ref{resLdL1}) reads
\begin{align}
\null &
\frac{1}{\beta}\,\bigg[A_{1}^{N-2(\mathsf{k}_1+1)}
\bigg(\frac{N-1}{2}-\mathsf{k}_1-\sum_{j=2}^{N}\frac{A_1}{A_1- A_j} 
\bigg) \prod_{i=2}^{N}\frac{1}{(A_1 - A_i)^2}
\bigg] (\beta\Lambda_1)^{2N} 
\\
& \hspace{0.4cm} +
\frac{1}{\beta}\,\bigg[ A_{1}^{N-2(\kfive-\mathsf{k}_1+1)}
\bigg( \frac{N-1}{2}-\kfive+\mathsf{k}_1-\sum_{j=2}^{N}\frac{A_1}{A_{1j}} \bigg)
\prod_{i=2}^{N} 
\frac{1}{(A_1-A_i)^2} \bigg] \frac{(\beta\Lambda)^{4N}}{(\beta\Lambda_1)^{2N}}~,
\nonumber
\end{align}
and agrees with the two-instanton term  of the logarithmic $q_1$-derivative of the superpotential
${\mathcal W}^{\,(\text{I})}_{\text{inst}}$ computed using localization methods 
with the first JK prescription described in Section~\ref{localization},
provided $\vert\,\kfive\,\vert < N$.

We have made numerous checks at higher instanton numbers and for various values of $N$, always 
finding a perfect match between localization and chiral ring analysis provided the relations 
(\ref{q1q2Lambda}) and (\ref{m1m2k}) are used. 

\subsubsection*{Generic surface operators}
\label{subsec:gencase}
We now consider a generic surface operator. In order to test the correspondence between the solution of 
the chiral ring equations and the localization results, it is crucial to connect the parameters used in 
the two descriptions and generalize (\ref{q1q2Lambda}) and (\ref{m1m2k}).
To this purpose, it is useful to recall that in deriving these relations it is was sufficient to consider the 
1-instanton result. Moreover, in comparing \eqref{resLdL1} and the $q_1$-logarithmic derivative of the
superpotential (\ref{W1instsimpleJK1}), we kept fixed the scale of the 5d theory. If we temporarily 
set $\Lambda=0$, and thus freeze the 5d dynamics, it becomes feasible to explicitly compute
the 1-instanton contribution to the solution of the chiral ring equations for a generic surface operator
and then compare with the localization results. Once this is done, it is possible to reinstate the 
dependence on $\Lambda$ in a rather simple manner, and find the generalization of the maps
(\ref{q1q2Lambda}) and (\ref{m1m2k}). 
Since the derivation is a bit lengthy, we discuss it in Appendix~\ref{appgenericsurfop}. Here
we simply report the final result which is quite simple:
\begin{equation}
\begin{aligned}
\label{genmapqm}
q_I &=  -(-1)^{r_I}~(\beta\Lambda_I)^{b_I}\qquad\text{for}
\quad I = 1, \ldots,  M-1~,\\
q_M&=(-1)^N\,(\beta\Lambda)^{2N}\Big(\prod_{I=1}^{M-1} q_I\Big)^{-1}~,
\end{aligned}
\end{equation}
and
\begin{equation}
\begin{aligned}
\label{mIgenare}
\mathsf{m}_I &= \mathsf{k}_I\qquad\text{for} \quad I = 1, \ldots,  M-1~,\\
\mathsf{m}_M&= \kfive- \sum_{I=1}^{M-1} \mathsf{k}_I~.
\end{aligned}
\end{equation}
Using these maps, we have investigated many different surface operators at the first few instanton orders
and found that the relation 
\begin{equation}
\label{genreltw}
\frac{1}{b_I}\,\Lambda_I\frac{d\cW_\star}{d\Lambda_I}
= \frac{1}{\beta}\,\tr \log S^{(I)}_{\star,\mathrm{class}} 
+ q_I \frac{d{\mathcal W}^{\,(\text{I})}_{\text{inst}}}{dq_I}~,
\end{equation}
which generalizes (\ref{logderqWs}), is always obeyed if $\vert\,\kfive\,\vert< N$.

\subsection{The dual linear quiver and its twisted chiral ring equations}
We now address the question of whether it is possible to establish a connection between the chiral 
ring equations and the localization results for the other JK prescription. 
This analysis will allow us to clarify the map between different residue prescriptions and 
distinct quiver realizations of the same surface operator. 
 
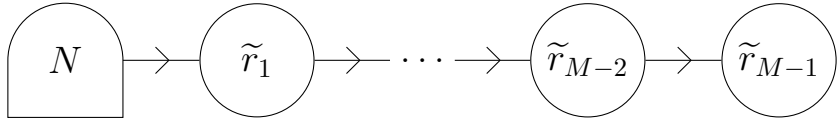
\begin{figure}[ht]
\begin{center}
\begin{tikzpicture}[decoration={
markings,mark=at position 0.6 with {\draw (-5pt,-5pt) -- (0pt,0pt);
                \draw (-5pt,5pt) -- (0pt,0pt);}}]
  \matrix[row sep=10mm,column sep=5mm] {
       \node(gfN)[gaugedflavor]{\Large $N$}; & & 
       \node(g2)[gauge] {\Large $\widetilde{r}_{1}$};  && 
       \node(dots){\Large $\ldots$}; & & 
       \node(glast)[gauge] {\Large $\widetilde{r}_{M-2}$};& &
      \node(g1)[gauge] {\Large $\widetilde{r}_{M-1}$}; \\
  };
\graph{(gfN) --[postaction={decorate}](g2) --[postaction={decorate}](dots)--[postaction={decorate}](glast) 
--[postaction={decorate}](g1);};
\end{tikzpicture}
\end{center}
\vspace{-0.5cm}
\caption{The quiver theory which is dual to the one in Fig.~\ref{quiverpicgeneric1}.}
\label{dualquiverpicgeneric1}
\end{figure}

Building on the results of \cite{Ashok:2017lko}, we propose that the quiver theory that is 
relevant to match the localization prescription with the JK parameter $\widetilde{\eta}$ given 
in (\ref{JKDual}) is the one represented in Fig.~\ref{dualquiverpicgeneric1}.
Here the ranks $\widetilde{r}_I$ of the 3d gauge groups are related to the partition 
$\vec{n}=[n_1, n_2, \ldots \, , n_M]$  that labels the surface operator according to
\cite{Gorsky:2017hro, Ashok:2017lko}
\begin{equation}
\widetilde{r}_{I} = N- \sum_{J=1}^{I} n_J=N-r_I~.
\label{tilderI} 
\end{equation}
As for the original quiver of Fig.~\ref{quiverpicgeneric1}, in the present case 
the low-energy effective theory on the Coulomb moduli space is parameterized by 
the diagonal components of the complex scalar fields in the adjoint twisted chiral multiplets, which
we denote as
\begin{equation}
\widetilde{\sigma}^{(I)}=\text{diag}\big\{\,\widetilde{\sigma}^{(I)}_1,\widetilde{\sigma}^{(I)}_2,
\ldots,\widetilde{\sigma}^{(I)}_{\widetilde{r}_I}\,\big\}
\end{equation}
for $I=1,\cdots,M-1$. The twisted chiral superpotential corresponding to this quiver 
takes the form
\begin{align}
\widetilde{W}=&\sum_{I=1}^{M-1}\sum_{s=1}^{\widetilde{r}_I}
\widetilde{b}_I\,\log (\beta\widetilde{\Lambda}_I)\,\widetilde\sigma^{(I)}_s
- \sum_{I=2}^{M-1} \sum_{s=1}^{\widetilde{r}_{I}} \sum_{t=1}^{\widetilde{r}_{I-1}}
\ell\big(\widetilde\sigma^{(I-1)}_t - \widetilde\sigma^{(I)}_s\big)-\sum_{s=1}^{\widetilde{r}_1}\Big\langle 
\Tr \ell\big(\Phi-\widetilde\sigma^{(1)}_s \big) \Big\rangle
\notag\\
&\qquad
-\sum_{I=1}^{M-1}\frac{\beta\,\widetilde{\mathsf{k}}_I }{2}\,\Tr \big(\widetilde{\sigma}^{(I)}\big)^2
\label{dtildeW15d}
\end{align}
where $\widetilde{\Lambda}_I$ is the (complexified) strong-coupling scale of the $I$-th node and 
the last term accounts for the Chern-Simons interactions on the 3d nodes with couplings 
$\widetilde{\mathsf{k}}_I $. 
The parameters $\widetilde b_I$ are defined as\,\footnote{Here and in the
following we understand that $\widetilde{r}_0=N$ and $\widetilde{r}_M=0$.}
\begin{equation}
\widetilde b_I = \widetilde{r}_{I+1}-\widetilde{r}_{I-1}
\end{equation}
and, because of (\ref{tilderI}), they are related to the analogous 
parameters $b_I$ introduced in the earlier quiver as:
\begin{equation}
\widetilde b_I = -b_I~.
\label{btildeb}
\end{equation}
The chiral ring equations, obtained by extremizing $\widetilde{\mathcal{W}}$, can be 
concisely expressed in terms of the functions
\begin{equation}
\widetilde{Q}_I(Z)= Z^{-\frac{\widetilde{r}_I}{2}}
 \prod_{s=1}^{\widetilde{r}_I}\big(\widetilde{S}_s^{(I)}\big)^{-\frac{1}{2}}\,\big(Z-
 \widetilde{S}_s^{(I)}\big)
 \label{tildeQIZ}
\end{equation}
where $\widetilde{S}_s^{(I)}=\rme^{\,\beta\,\widetilde{\sigma}^{(I)}_s}$ in complete analogy with (\ref{QIZ}).
Indeed, for $I=2,\cdots,M-1$ we find
\begin{equation}
\widetilde{Q}_{I-1}(Z)
= (-1)^{\widetilde{r}_{I-1}}\,(\beta\widetilde{\Lambda}_I)^{-\widetilde{b}_{I}} \,
Z^{\widetilde{\mathsf{k}}_{I}}\, \widetilde{Q}_{I+1}(Z)~,
\end{equation}
for $Z=\widetilde{S}_s^{(I)}$, while the chiral ring equation of the first node ($I=1$)
involves the resolvent of the 5d SU($N$) theory and reads
\begin{equation}
P_N(Z) 
= (-1)^N(\beta\widetilde{\Lambda}_1)^{-\widetilde{b}_{1}}\, 
Z^{\widetilde{\mathsf{k}}_{1}}\, \widetilde{Q}_2(Z)
+ (-1)^N\frac{(\beta\Lambda)^{2N}}{(\beta\widetilde{\Lambda}_1\phantom{})^{-\widetilde{b}_{1}}}\,
\frac{Z^{-\kfive-\widetilde{\mathsf{k}}_1 }}{ \widetilde{Q}_2(z)}
\label{dualCReq}
\end{equation}
for $Z= \widetilde{S}_s^{(1)}$.

The classical vacuum around which we perturbatively solve the above equations is
\begin{equation}
\label{ansd}
\widetilde{S}^{(I)}_{\star,\mathrm{class}} 
= \diag(A_{N-\widetilde{r}_{I} +1}, \ldots, A_{N})~.
\end{equation}
This corresponds to simply choosing for each node $I$ 
the complement set of $A_u$ that appear in the classical vacuum (\ref{Scl}) 
for the corresponding node in the original quiver. 
Using an ansatz analogous to the one in (\ref{ansS}) and expanding in 
powers of $\beta$ around (\ref{ansd}),
we can obtain the solution $\widetilde{S}^{(I)}_{\star}$ of the chiral ring equations 
to the desired perturbative order and, in analogy with (\ref{logLIder}), relate it to
the logarithmic derivative with respect to $\widetilde{\Lambda}_I$ of the twisted superpotential 
evaluated on this solution, namely
\begin{equation}
\label{logLIderdual}
\widetilde{\Lambda}_I\frac{d\widetilde{\cW}_\star}{d\widetilde{\Lambda}_I}
= \frac{\widetilde{b}_I}{\beta}\,\tr \log \widetilde{S}^{(I)}_{\star} ~.
\end{equation}
We now give some details in the case of the simple operators of type $[1,N-1]$.
 
\subsubsection*{Simple surface operators}
\label{1N1dual}
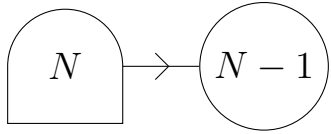
\begin{figure}[ht]
\begin{center}
\begin{tikzpicture}[decoration={
markings,mark=at position 0.6 with {\draw (0pt,0pt) -- (-5pt,-5pt);
                \draw (0pt,0pt) -- (-5pt,5pt);}}]
  \matrix[row sep=10mm,column sep=5mm] {
       \node(gf2)[gaugedflavor]{\Large $N$};    & &\node(g1)[gauge] {\Large $N-1$}; \\
  };
\graph{(gf2) --[postaction={decorate}](g1);};
\end{tikzpicture}
\end{center}
\vspace{-0.5cm}
\caption{The dual quiver for the $[1,N-1]$ defect in the SU($N$) theory.}
\label{fig:su312dq}
\end{figure}

In this case the quiver has a single 3d gauge node and is as depicted in Fig.~\ref{fig:su312dq}.
Correspondingly, using $\widetilde{b}_1=-N$ and $\widetilde{Q}_2=1$, we see that 
the twisted chiral ring equations (\ref{dualCReq}) take the following form:
\begin{equation}
P_N(Z) =(-1)^N(\beta\widetilde{\Lambda}_1)^{N}\,
Z^{\widetilde{\mathsf{k}}_{1}}
+(-1)^N\frac{(\beta\Lambda)^{2N}}{(\beta\widetilde{\Lambda}_1\phantom{})^{N}}\, Z^{-\kfive-\widetilde{\mathsf{k}}_1}
\end{equation}
for $Z= \widetilde{S}_s^{(1)}$ with $s=1,\cdots,N-1$. To leading order these equations are solved by
\begin{equation}
\begin{aligned}
\widetilde{S}_{\star,s}^{(1)}&= A_{s+1} \bigg[1+(-1)^N \Big(
 (\beta\widetilde{\Lambda}_1)^{N} A_{s+1}^{\frac{N}{2}+\widetilde{\mathsf{k}}_1-1}\!+ 
 \frac{(\beta\Lambda)^{2N}}{(\beta\widetilde{\Lambda}_1\phantom{})^{N}}\, 
 A_{s+1}^{\frac{N}{2}-\kfive-\widetilde{\mathsf{k}}_1-1}\Big)\\
& \phantom{ = }~ ~~
\times \prod_{\substack{j=1\\\ j\ne s+1}}^N\!\!
\frac{1}{(A_{s+1}- A_j)}  +\ldots \bigg]~.
\end{aligned}
\end{equation}
Exploiting (\ref{logLIderdual}), we get
\begin{align}
\frac{1}{N}\,\widetilde{\Lambda}_1\frac{d{\widetilde{\mathcal W}_{\star}}}{d\widetilde{\Lambda}_1}
& =-\frac{1}{\beta}\,\tr\,\log \widetilde{S}^{(1)}_\star\phantom{\Big|}\nonumber\\
& = -\frac{1}{\beta}\,\sum_{i=2}^N\log A_i-\frac{1}{\beta}\sum_{i=2}^N\bigg[
(-1)^N\Big(
\!(\beta\widetilde{\Lambda}_1)^{N} A_{i}^{\frac{N}{2}+\widetilde{\mathsf{k}}_1-1}\!\!+ 
 \frac{(\beta\Lambda)^{2N}}{(\beta\widetilde{\Lambda}_1\phantom{})^{N}}\, 
 A_{i}^{\frac{N}{2}-\kfive-\widetilde{\mathsf{k}}_1-1}
 \Big)\nonumber\\
 &
\phantom{ = }~ ~~\times\prod_{\substack{j=1\\ j\ne i}}^N
 \frac{1}{(A_i- A_j)}\bigg]
+ \ldots ~.\label{resLdL1dual}
\end{align}
The quantity in square brackets has the same structure of the (logarithmic derivative of the) twisted
superpotential (\ref{W1instsimpleJK2}) computed using the second JK prescription.
Indeed, if make the following identifications
\begin{equation}
q_1=(-1)^N (\beta\widetilde{\Lambda}_1)^{N} ~,\qquad
q_2=\frac{(\beta\Lambda)^{2N}}{(\beta\widetilde{\Lambda}_1)^N}~,
\label{q1q2tildeL}
\end{equation}
and
\begin{equation}
\mathsf{m}_1 = -\widetilde{\mathsf{k}}_1~,\qquad
\mathsf{m}_{2}=\kfive +\widetilde{\mathsf{k}}_1~,
\label{dualm1m2k}
\end{equation}
we obtain
\begin{equation}
\label{logderqWdual}
\frac{1}{N}\,{\widetilde\Lambda}_1\frac{d{\widetilde{\mathcal W}}_{\star}}{d\widetilde{\Lambda}_1} 
 = -\frac{1}{\beta}\sum_{i=2}^N \log A_i 
+  q_1 \frac{d{\widetilde{\mathcal W}}^{\,(\text{II})}_{1-\text{inst}}}{dq_1}+ \cdots~,
\end{equation}
where on the right hand side the derivative with respect to $q_1$ is taken by keeping the product 
$q_1 q_2$ fixed. This is clearly the counterpart in the dual quiver of the relation (\ref{logderqWs})
that we found in the original theory. Notice that the identifications (\ref{q1q2Lambda}) and (\ref{m1m2k})
continue to hold, but the map between the localization parameter $\mathsf{m}_1$ and the 3d
Chern-Simons coupling has an opposite sign as compared to the original quiver. We have checked in several
examples that the higher-instanton corrections to the left hand side of (\ref{logderqWdual}) fully agree
with those of the logarithmic derivative of ${\mathcal W}^{\,(\text{II})}_{\text{inst}}$, computed 
using localization with the second JK prescription, provided $\vert\,\kfive\,\vert < N$.

\subsubsection*{Generic surface operators}
The above procedure can be applied to a generic surface operator. The details are given in 
Appendix~\ref{appgenericsurfop}. Here we merely report the maps between the parameters
used in the localization calculations and those appearing in the chiral ring equations:
\begin{equation}
\begin{aligned}
\label{genmapqmdual}
q_I &=  -(-1)^{\widetilde{r}_I}~(\beta\widetilde{\Lambda}_I)^{-\widetilde{b}_I}\qquad\text{for}
\quad I = 1, \ldots,  M-1~,\\
q_M&=(-1)^N\,(\beta\Lambda)^{2N}\Big(\prod_{I=1}^{M-1} q_I\Big)^{-1}~,
\end{aligned}
\end{equation}
and
\begin{equation}
\begin{aligned}
\label{mIgenaredual}
\mathsf{m}_I &= -\widetilde{\mathsf{k}}_I\qquad\text{for} \quad I = 1, \ldots,  M-1~,\\
\mathsf{m}_M&= \kfive+ \sum_{I=1}^{M-1} \widetilde{\mathsf{k}}_I~.
\end{aligned}
\end{equation}
Using these maps, we have checked in several examples at the first few instanton orders
that the relation 
\begin{equation}
\label{genreltwdual}
\frac{1}{\widetilde{b}_I}\,\widetilde{\Lambda}_I\frac{d\widetilde{\cW}_\star}{d\widetilde{\Lambda}_I}
= \frac{1}{\beta}\,\tr \log S^{(I)}_{\star,\mathrm{class}} 
- q_I \frac{d{\mathcal W}^{\,(\text{II})}_{\text{inst}}}{dq_I}~,
\end{equation}
which generalizes (\ref{logderqWdual}) to a generic surface operator, is always satisfied provided 
$\vert\kfive\vert < N$.

\subsection{Summary}
We have discussed in detail how two different realizations of a surface defect encoded in 
the two quiver diagrams of Fig.~\ref{quiverpicgeneric1} and Fig.~\ref{dualquiverpicgeneric1} correspond, 
respectively,
to the two different JK prescriptions used in the localization approach. We stress that the integrand in the
ramified instanton partition function remains the same, and in particular that the parameters $\mathsf{m}_I$
do not change;
what changes is the map between these parameters and the Chern-Simons coefficients
of the 3d nodes in the two quiver theories.
Our results can be summarized in the following diagram.
\begin{equation}
\begin{tikzpicture}[grow via three points={one child at (0,-1) and two children at (-2.2,-2.2) 
and (3,-4.3)},decoration={
markings,
mark=at position 0.6 with {\draw (-1ex,-1ex) -- (0ex,0ex);
                \draw (-1ex,1ex) -- (0ex,0ex);}}]
\node at (-0.5,0) {$\displaystyle{\prod_{I=1}^M \bigg[\frac{(-q_I)^{d_I}}{d_I!}\,
\int_{\mathcal{C}} \,\prod_{\sigma=1}^{d_I} 
\Big(\beta\,\frac{d\chi_{I,\sigma}}{2\pi\ii}\Big)\, 
\rme^{-\beta\,\mathsf{m}_I \chi_{I,\sigma} } \bigg]~
z_{\{d_I\}}}$} 
child  {
  edge from parent
        node[left] {$\mathrm{JK}_{\mathrm{I}}\ $}
  }
child {
  edge from parent
        node[right] {$\ \mathrm{JK}_{\mathrm{II}}$}
  }
;\\
\small \matrix[row sep=2ex,column sep=0.2em] at (0,-4.3){
      \node(g1)[gaugeS,label=-70:$\mathsf{m}_1$] {$\phantom{\Big|}r_1\phantom{\Big|}$};   
      & \node(g2)[gaugeS,label=-70:$\mathsf{m}_2$] {$\phantom{\Big|}r_2\phantom{\Big|}$};  
      & \node(dots)[label=-70:$\phantom{x}$]{$\ldots$};
       && \node(glast)[gaugeS,label=-70:$\mathsf{m}_{M-1}$] {\small{$r_{M-1}$}}; 
       &\node(fN)[gaugedflavorS,label=-50:$\kfive$]{$~N~$};
       &&&
       && \\
         & &
       &&  &\node(ffN)[gaugedflavorS,label=-50:$\kfive$]{$~N~$};
       & \node(gg1)[gaugeS,label=-70:$-\mathsf{m}_1$] {$\phantom{\Big|}\tilde{r}_1\phantom{\Big|}$};   
       & \node(gg2)[gaugeS,label=-70:$-\mathsf{m}_2$] {$\phantom{\Big|}\tilde{r}_2\phantom{\Big|}$};  
       & \node(ddots)[label=-70:$\phantom{x}$]{$\ldots$};
       && \node(gglast)[gaugeS,label=-80:$-\mathsf{m}_{M-1}$] {\small{$\tilde{r}_{M-1}$}};\\
  };
\graph{(g1) --[postaction={decorate}](g2) --[postaction={decorate}](dots)--[postaction={decorate}](glast) 
--[postaction={decorate}](fN);};
\graph{(ffN)--[postaction={decorate}](gg1) --[postaction={decorate}](gg2) --[postaction={decorate}](ddots)--[postaction={decorate}](gglast);};
\end{tikzpicture}
\end{equation}
In the next section we discuss how the two quiver theories are related to each other by
IR Aharony-Seiberg dualities.

\section{Relation to Aharony-Seiberg dualities}
\label{secn:AharonySeiberg}

In Section~\ref{localizecalc}, we studied surface operators realized as Gukov-Witten defects
by means of localization techniques and computed the ramified instanton partition function from 
which the twisted chiral superpotential can be extracted. 
Besides the instanton counting parameters $q_I$, our results
depend on the parameters $\mathsf{m}_I$ that were introduced
as counterparts of the Chern-Simons couplings that may appear when the surface defects
are represented as coupled 3d/5d systems.
Localization requires a residue prescription, usually specified by means of a Jeffrey-Kirwan parameter, in
order to select the poles contributing to the integral over the instanton moduli space. 
We have computed the twisted superpotential using two different (and complementary) prescriptions and 
shown that only when the parameters $\mathsf{m}_I$ satisfy the constraints (\ref{cond1mI}) 
and (\ref{cond2mI}) the two results agree.

On the other hand, in Section~\ref{secn:3d5d}, we considered the realization of the defect 
by means of two different coupled 3d/5d quiver theories. They give rise to twisted chiral 
superpotentials that exactly match those arising from the two localization residue prescriptions, 
provided the parameters $m_I$ are mapped to the 3d and 5d Chern-Simons levels $\mathsf{k}_I$ and 
$\kfive$ according to (\ref{mIgenare}) or (\ref{mIgenaredual}). 
Therefore, the conditions on $\mathsf{m}_I$ under which the two localization prescriptions yield the 
same result must correspond to the conditions that the Chern-Simons parameters must obey 
in order for the two quiver theories to be dual to each other.
In the following, we explore the physical content of these constraints and their connection with related work
in the literature.

Let us first consider the quiver theory of Fig.~\ref{quiverpicgeneric1}, and for simplicity turn off
the 5d dynamics on the rightmost node in order to have a purely 3d theory. 
This corresponds to setting the 5d scale $\Lambda$ to zero and to considering SU($N$) as a 
global flavour 
symmetry\,\footnote{From the localization point of view, setting the 5d scale to zero reduces the ramified instanton partition function to a 3d vortex partition function.}. 
For $I = 1,\ldots, M-1$ the constraints (\ref{cond1mI}) and (\ref{cond2mI}) become 
\begin{equation}
\label{CSlevelconst1}
 \mathsf{k}_I + \frac{b_I}{2}\in \mathbb{Z}~,
\end{equation} 
and
\begin{equation}
\label{CSlevelconst2}
 \vert\, \mathsf{k}_I\,\vert \leq \frac{b_I}{2} - 1~.
\end{equation}
Here we have used the fact that $b_I = r_{I+1}-r_{I-1}$
and, as before, understood that $r_0=0$ and $r_M=N$. 
These constraints and their physical interpretation are well known. The integrality condition 
(\ref{CSlevelconst1}) is a requirement on the absence of the $\mathbb{Z}_2$ parity anomaly 
in three dimensions \cite{Redlich:1983kn, Redlich:1983dv}, and is related to the fact that integrating out 
an odd number of chiral fermions leads to a half-integer Chern-Simons term at one-loop. Indeed,
$b_I$ is the effective number of chiral (fundamental) matter at the $I$-th node.
The inequality (\ref{CSlevelconst2}) is the constraint found in \cite{Benini:2011mf} 
(see in particular Eq.~(3.51) of this reference) for the so-called ``maximally chiral 
theories''. Notice that the 3d gauge theory at each node of the quiver belongs 
to this class, since the ranks $r_I$ are monotonically increasing with $I$.

When the constraint (\ref{CSlevelconst2}) is satisfied, the $U(r_I)_{\mathsf{k}_I}$ theory 
at the $I$-th node admits an Aharony-Seiberg dual, which is a $U(r_{I+1}\!-\!r_I)_{-\mathsf{k}_I}$ 
theory\,\footnote{In more general situations, the dual rank is $\max(s,s^\prime)- r_I$, 
where $s$ and $s^\prime$ are the numbers of chiral and anti-chiral matter multiplets charged 
with respect to the $I$-th gauge group.} possessing additional (mesonic) fields 
with a superpotential term \cite{Seiberg:1994pq, Aharony:1997gp, Aharony:2014uya}.
By performing subsequent duality transformations, one may obtain 
many distinct dual quiver theories. In particular, one can check that by successively applying 
such dualities to the quiver of Fig.~\ref{quiverpicgeneric1}, starting from the node 
with $I=M-1$ and proceeding all the way to the left-most node with $I=1$, 
one ends up with precisely the linear quiver of Fig.~\ref{dualquiverpicgeneric1} without any additional mesonic 
fields and superpotential terms. In fact, with the first duality transformation 
the $U(r_{M-1})_{\mathsf{k}_{M-1}}$ node becomes a $U(N\!-\!r_{M-1})_{-\mathsf{k}_{M-1}}$ theory
with mesons that behave as $N$ multiplets in the fundamental of $U(r_{M-2})_{\mathsf{k}_{M-2}}$.
Dualizing the latter, we obtain a $U(N\!-\!r_{M-2})_{-\mathsf{k}_{M-2}}$ theory along 
with $N$ mesons that transform in the fundamental of $U(r_{M-3})_{\mathsf{k}_{M-3}}$. 
Continuing in this dualization process, all superpotential terms cancel and we obtain the linear quiver of Fig.~\ref{dualquiverpicgeneric1}.

According to the analysis of \cite{Benini:2011mf}, theories with 3d Chern-Simons levels outside the range (\ref{CSlevelconst2}) still admit Aharony-Seiberg duals, but the ranks of the gauge groups for the latter depend on the Chern-Simons levels and in certain cases exceed the 
rank of the global flavour symmetry. If this is the case, turning on twisted masses for the flavours would not completely lift the Coulomb branch and the resulting 3d low-energy effective theory is not massive. 
Thus, these dual models cannot represent Gukov-Witten defects in a higher dimensional theory since 
the general picture of surface operators as coupled gauge theories proposed in \cite{Gaiotto:2009fs} 
necessarily assumes the fibration of a discrete set of vacua, namely the solutions to the twisted chiral rings of 
the lower dimensional theory, over the Coulomb moduli space of the higher dimensional theory.

Let us now consider the case when the five dimensional gauge coupling is turned on. From the localization point of view, we now have to take into account the case $I=M$ in (\ref{cond1mI}) and (\ref{cond2mI}). This leads to the condition (\ref{k5dconstraintSO}) 
on the 5d Chern-Simons coupling, {\it{i.e.}}
\begin{equation}
\label{k5dmodifiedSO}
\vert\,\kfive\,\vert\leq N-M ~.
\end{equation}
The same bound can be derived from the twisted chiral ring relations.
Consider for simplicity the surface operator of type $[1,N-1]$, corresponding to $M=2$, for which
the twisted chiral ring equation (see \eqref{PNeqnv1}) is
\begin{align}
Z^{\frac N2} P_N(Z) & \equiv 
Z^N + \sum_{i=1}^{N-1} (-1)^i \,Z^{N-i}\, U_{N-i}(\kfive) + (-1)^N
\nonumber\\
& =(\beta\Lambda_1)^{N}
Z^{\frac{N}{2}-\mathsf{k}_1} + \frac{(\beta\Lambda)^{2N}}{(\beta\Lambda_1)^{N}}Z^{\frac{N}{2}-
\kfive + \mathsf{k}_{1}}
~,
\label{CRfin}
\end{align}
with $Z=S^{(1)}$. In our analysis we assumed that it was possible to find a solution of this equation 
as a power series expansion around a classical vacuum specified by
$S^{(1)}_{\star,\mathrm{class}}= \rme^{\beta\,a_u}$, with $a_u$ being one of the $N$ vacuum 
expectation values of the 5d adjoint scalar $\Phi$. Following the discussion in \cite{Gaiotto:2009fs} for the (dimensionally reduced) 2d/4d case, one can analyze the fibration of these discrete solutions over the moduli space of the higher dimensional gauge theory and, from the geometry of the total space, recover the form of the Seiberg-Witten curve of the compactified 5d theory. This can be seen by defining \cite{Brini:2008rh}
\begin{equation}
Y =
(\beta\Lambda_1)^{N}Z^{-\mathsf{k}_1} - \frac{(\beta\Lambda)^{2N}}{(\beta\Lambda_1)^{N}}Z^{-
\kfive + \mathsf{k}_{1}}~,
\end{equation}
and noting that from (\ref{CRfin}) we have $Y^2 = P_N^2 - 4(\beta\Lambda)^{2N} Z^{-\kfive}$. 

However, the chiral ring equations are related to the twisted superpotential that arises in presence of the 
defect, and contain more information than the Seiberg-Witten curve, which encodes the prepotential of the 
pure 5d theory. It is easy to check that demanding (\ref{CRfin}) to be 
a monic polynomial in $Z$ of degree $N$, whose constant term is set 
to be $(-1)^N$ by the SU$(N)$ condition, implies, beside
the conditions (\ref{CSlevelconst1}) and (\ref{CSlevelconst2}), also the 
relation $\vert\,\kfive\,\vert\leq N-2$, which is the bound (\ref{k5dmodifiedSO}) for $M=2$. 
The same kind of analysis in the case with no defect, \emph{i.e.} $M=1$, 
leads to the standard relation $\vert\,\kfive\,\vert \leq N-1$ in agreement with 
\cite{Intriligator:1997pq} (see the discussion after (\ref{PNclass})).

We pause to remark that for $|\kfive| < N$, there is perfect agreement between the twisted chiral 
superpotentials calculated using localization and the chiral ring analysis. Thus in this range of the 5d 
Chern-Simons level, one can study the surface operator either as a monodromy defect or as 
a coupled 3d/5d system. However, what the constraint (\ref{k5dmodifiedSO}) implies is that, 
for $N-M<\vert\,\kfive\,\vert < N$, due to a non vanishing contribution from the residue at zero 
or infinity, the superpotentials calculated using the two contour prescriptions differ. 
It is possible that in this range of $\kfive$ one might need to modify the contour integral description
of the defect and/or take into account extra light degrees of freedom in order to relate the two 
contour prescriptions. 
On the quiver side, this would require a more detailed understanding of Aharony-Seiberg dual theories. It 
would be very interesting to explore these possibilities.

In this work we have focused  on the two linear quivers at the end of a chain of duality transformations. 
It would be nice to better understand the twisted chiral rings and the 
superpotentials of the intermediate quivers obtained along the way. It would also be important to understand the map between such 3d/5d theories and the different residue 
prescriptions that can be considered in the localization integral. We leave these issues to future work.

\vskip 1.5cm
\noindent {\large {\bf Acknowledgments}}
\vskip 0.2cm
We thank Sourav Ballav, Madhusudhan Raman, Sujoy Mahato, and especially Amihay Hanany for many useful discussions. 

\noindent
The work of M.B.,~M.F.,~R.R.J.~and~A.L. is partially supported by the MIUR PRIN Contract 
2015MP2CX4 ``Non-perturbative Aspects Of Gauge Theories And Strings''.
\vskip 1cm
\begin{appendix}
\section{Chiral correlators in 5d gauge theories}
\label{chiralcorr}

In this appendix we outline the method of calculating the quantum chiral correlators in a
5d gauge theory, generalizing the discussion in \cite{Ashok:2017lko} to include 
a non-zero Chern-Simons coupling. The key idea is to start from the formula for the chiral correlators 
in 4d theories \cite{Bruzzo:2002xf, Losev:2003py,  Flume:2004rp,Ashok:2016ewb},
and suitably generalize it to the 5d case, namely
\begin{equation}
\label{mastervevs}
V_{\ell} = \Big\langle \Tr \rme^{\ell \beta \Phi}\Big\rangle
= \sum_{u=1}^N A_u^{ \ell} - \frac{1}{Z_{\text{inst}}}\,\sum_{k=1}^\infty  
\frac{q^k}{k!} \int_{\mathcal{C}} \bigg(\prod_{\sigma=1}^k \frac{d\chi_\sigma}{2\pi\ii}\bigg)
\, z_k(\chi_\sigma)\, {\mathcal O}_\ell(\chi_\sigma)~.
\end{equation}
Here, $Z_{\text{inst}}$ is the instanton partition function defined in (\ref{ZY5d}), $z_k(\chi_\sigma)$ is 
the integrand (\ref{5dCSintegrand}), and ${\mathcal O}_\ell$ is the following combination 
\cite{Ashok:2017lko}
\begin{equation}
{\mathcal O}_{\ell}(\chi_\sigma) = \sum_{\sigma=1}^k \rme^{\,\ell \,\beta\,\chi_\sigma} 
\big(1-\rme^{\,\ell\, \beta\,\epsilon_1}\big)\big(1-\rme^{\,\ell\,\beta\,\epsilon_2}\big) ~.
\end{equation}
At the 1-instanton level, by performing explicitly the integral over $\chi$ in (\ref{mastervevs}) we find
\begin{equation}
\label{Vellgeneral}
V_{\ell} 
= \sum_{u=1}^N A_{u}^{\ell} + \ell^2 (\beta\Lambda)^{2N}\sum_{u=1}^N \Bigg[
\frac{A_u^{N-2+\ell-\kfive}}{\displaystyle{\prod_{u\ne v}(A_{u}-A_{v})^2}}\Bigg] 
+ O\left((\beta\Lambda)^{4N}\right)~.
\end{equation}
The generating function for the $V_{\ell}$ is the resolvent of the SU($N$) gauge theory:
\begin{equation}
\label{Vdefn}
{T}=  N + 2\sum_{\ell} \frac{V_{\ell}}{Z^{\ell}}
\end{equation}
for which in Section~\ref{secn:5dCS} we proposed an explicit formula in terms of the
functions appearing in the Seiberg-Witten curve of the theory (see (\ref{5dresolventk})).
Working out the large $Z$ expansion, we obtain
\begin{align}
\label{Tzexp}
T=&~ N + 2\,\frac{U_1(\kfive)\!+\!(\beta\Lambda)^{2N}\delta_{\,\kfive,1-N}}{Z} 
\notag\\
&+2\, \frac{U_1^2(\kfive)\!-\!2\,U_2(\kfive)
\!+\!\big(4(\beta\Lambda)^{2N}U_1(\kfive)\!+\!3(\beta\Lambda)^{4N}\big)\delta_{\,\kfive,1-N}
\!+\!2(\beta\Lambda)^{2N}
\delta_{\,\kfive,2-N}}{Z^2} \notag
\\&+O(Z^{-3})
\end{align}
where $U_i(\kfive)$ are the gauge invariant coordinates on the Coulomb branch of the 5d
theory with Chern-Simons coupling $\kfive$. Comparing with (\ref{Vdefn}), we deduce
\begin{equation}
\begin{aligned}
U_1(\kfive)&=V_1-(\beta\Lambda)^{2N}\delta_{\,\kfive,1-N}~,\\
U_2(\kfive)&=\frac{1}{2}\big(V_1^2-V_2\big)+(\beta\Lambda)^{2N}\big(V_1\,\delta_{\,\kfive,1-N}+
\delta_{\,\kfive,2-N}\big)~.
\end{aligned}
\label{U1U2k}
\end{equation}
Similar formulas can be easily worked out for higher $U_i(\kfive)$ without any difficulty. 
However, since they are a bit involved we do not report them here.
Instead, as an illustrative example, we consider the explicit expression of the above formulas 
in the case of the SU(3) theory for which $U_1(\kfive)$ and $U_2(\kfive)$ are the two
independent coordinates of the quantum Coulomb branch.
In this case, using (\ref{Vellgeneral}) into (\ref{U1U2k}) and taking into account the SU(3) condition, 
we find 
\begin{equation}
\begin{aligned}
U_1(\kfive)&=\sum_{u=1}^3A_u+(\beta\Lambda)^6\Bigg[\sum_{u=1}^3
\frac{A_u^{2-\kfive}}
{\displaystyle{\prod_{u\ne v\ne w}\!\!(A_{u}-A_{v})^2\,(A_{u}-A_{w})^2}}-\delta_{\,\kfive,-2}\Bigg]
+O\big((\beta\Lambda)^{12}\big)~,\\
U_2(\kfive)&=\!\!\sum_{u\ne v=1}^3A_uA_v+(\beta\Lambda)^6\Bigg[\sum_{u=1}^3
\frac{A_u^{-\kfive}}
{\displaystyle{\prod_{u\ne v\ne w}\!\!(A_{u}-A_{v})^2\,(A_{u}-A_{w})^2}}-\delta_{\,\kfive,2}\Bigg]
+O\big((\beta\Lambda)^{12}\big)~.
\end{aligned}
\end{equation}
{From} these expressions, one can check that 
\begin{equation}
\begin{aligned}
U_i(-\kfive)=
\widetilde U_{3-i} (\kfive)~,
\end{aligned}
\end{equation}
where $\widetilde{U}_i$ is obtained from $U_i$ through the inversion $A_u\to 1/A_u$.
This is a particular case of the relation (\ref{UUtilde}) discussed in Section~\ref{sectionResolvent}.
We have checked this relation also for groups of higher rank at the 1-instanton level, confirming its validity.

\section{Map of parameters for the generic surface operator}
\label{appgenericsurfop}

In this Appendix we consider a generic surface operator and calculate the 1-instanton contribution to its
twisted chiral superpotential using the two JK prescriptions discussed in the main text, with the purpose
of finding the map between the parameters introduced in the localization calculations and those appearing in
the quiver theory, focusing in particular on the 3d gauge nodes.

\subsection*{The twisted superpotential}
We start from the 1-instanton partition function 
$Z_{1-\text{inst.}}$. This is given in (\ref{Partition_function_gen_op2}), which we rewrite here for convenience
\begin{equation}
\begin{aligned}
\label{Partition_function_gen_opApp}
Z_{1-\text{inst}}&=- \sum_{I=1}^{M} {q_I}\,
\frac{E_1^{\frac {n_{I} + n_{I+1}}4+\frac{1}{2}}\, \hat{E}_2^{\frac {n_{I} + n_{I+1}}4}}{E_1 - 1}
\int_{\mathcal{C}} \frac{dX_I}{2\pi\ii} ~X_I^{\frac {n_{I} + n_{I+1}}2 - \mathsf{m}_I-1}\\
&\hspace{1.2cm}\times\prod_{\ell =r_{I-1}+1}^{r_I}
\frac{1}{\Big(A_{\ell}\, \sqrt{E_1\hat E_2\phantom{|}}-X_I\Big)}
 \times \prod_{j=r_{I}+1}^{r_{I+1}}\frac{1}{\Big(A_j -  X_I\,\sqrt{E_1\hat E_2\phantom{|}}\Big)} ~.
\end{aligned}
\end{equation}
Since our main goal is to find the 3d interpretation of the parameters, we can set $q_M=0$, which in 
view of (\ref{genmapqm}) and (\ref{genmapqmdual}) is equivalent to put $\Lambda=0$ and hence freeze out 
the quantum dynamical effects in the 5d theory. Then we remain only with the integrals over $X_I$ with $I=1,\cdots,M-1$.

In the JK$_{\text{I}}$ prescription only poles in the upper-half complex plane of $X_I$ are chosen. 
In our case they are
\begin{equation}
X_I = A_{\ell}\, \sqrt{E_1\hat E_2\phantom{|}}
\end{equation}
for $\ell = r_{I-1}+1, \ldots r_I$.
Evaluating the residues and extracting the twisted chiral superpotential according to (\ref{ExtractW}), 
we obtain
\begin{align}
{\mathcal W}^{\,(\text{I})}_{\text{1-inst}}
 =& -\frac{1}{\beta}\sum_{I=1}^{M-1}(-1)^{n_I}q_I
  \sum_{i=1}^{n_I} \Bigg[(A_{r_{I-1}+i})^{\frac{n_I +n_{ I + 1}}{2}  -\mathsf{m}_I - \frac{1}{2}}
  \prod_{\substack{j=1\\ j \neq i}}^{n_I}\,A_{r_{I-1}+j}^{\frac 12}
 \prod_{s=1}^{n_{I+1}} \, A_{r_{I} +s}^{\frac 12}\notag\\
 &\qquad\qquad\times\prod_{\substack{\ell=1\\ \ell \neq i}}^{n_I}\frac{1}{(A_{r_{I-1}+i} - A_{r_{I-1}+\ell})}
 \prod_{t=1}^{n_{I+1}}
 \frac{1}{(A_{r_{I-1}+i} - A_{r_{I+1}-t+1})}\Bigg]~.
 \label{WIgeneral}
\end{align}

With the JK$_{\text{II}}$ prescription, one makes the complementary choice of poles, namely those located at
\begin{equation}
X_I = \frac{A_{j}}{\sqrt{E_1\hat E_2\phantom{|}}}
\end{equation}
for $j= r_{I}+1, \ldots, r_{I+1}$. Computing the corresponding residues yields
\begin{align}
\label{WIIgeneral}
{\mathcal W}^{\,(\text{II})}_{\text{1-inst}}
 =& \frac{1}{\beta}\sum_{I=1}^{M-1}(-1)^{n_I}q_I
 \sum_{i=1}^{n_{I+1}}\Bigg[
 (A_{r_{I}+i})^{\frac{n_I+n_{ I + 1} }{2}  -\mathsf{m}_I- \frac{1}{2}} \prod_{j = 1}^{n_I}\,A_{r_{I-1}+j}^{\frac 12}
 \prod_{\substack{s=1\\ s \neq i}}^{n_{I+1}} \, A_{r_{I} +s}^{\frac 12} \notag
 \\
 &\qquad\qquad\times
 \prod_{\ell = 1}^{n_I} \frac{1}{(A_{r_{I}+i}-A_{r_{I-1}+\ell})}
 \prod_{\substack{t=1 \\ t \neq i}}^{n_{I+1}}\frac{1}{(A_{r_{I}+i}- A_{r_{I}+t})}\Bigg]~.
\end{align}
As we have seen in Section~\ref{subsecn:1inst}, these two expressions are in general different, 
unless the parameters $\mathsf{m}_I$ satisfy the conditions (\ref{cond1mI}) and (\ref{cond2mI}).

\subsubsection*{Linear quiver}

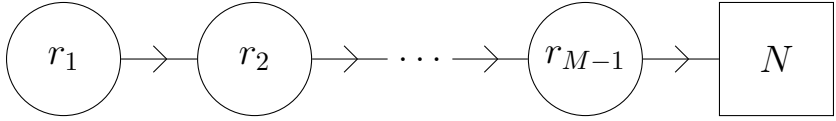
\begin{figure}[ht]
\begin{center}
\begin{tikzpicture}[decoration={markings, mark=at position 0.6 with {\draw (-5pt,-5pt) -- (0pt,0pt);       
         \draw (-5pt,5pt) -- (0pt,0pt);}}]
  \matrix[row sep=10mm,column sep=5mm] {
      \node (g1)[gauge] {\Large $r_1$};  & & \node(g2)[gauge] {\Large $r_2$};  && \node(dots){\Large 
      $\ldots$};
      & & \node(glast)[gauge] {\Large $r_{M-1}$};& &\node(gfN)[flavor]{\Large $N$};\\
  };
\graph{(g1) --[postaction={decorate}](g2) --[postaction={decorate}](dots)--[postaction={decorate}](glast) 
--[postaction={decorate}](gfN);};
\end{tikzpicture}
\end{center}
\vspace{-0.5cm}
\caption{The 3d quiver which lifts to the generic surface operator upon gauging the SU$(N)$ flavour 
symmetry represented by the square node at the right.}
\label{3dquiverpicgeneric1}
\end{figure}

We now study the twisted chiral ring equations whose solutions are the vacua of the 3d quiver
represented in Fig.~\ref{3dquiverpicgeneric1}.
Here the SU$(N)$ node on the right is not gauged, since our objective is to find the map between the 
3d parameters that include the Chern-Simons levels and the strong coupling scales of the gauge theory. 
In particular, this means that the 5d scale $\Lambda$ is set to 0, as we did before 
in the localization calculations.
All chiral ring equations for this quiver are given by
\begin{equation}
 Q_{I+1}(S_s^{(I)})= (-1)^{r_{I-1}} (\beta \Lambda_I )^{b_I} \big(S_s^{(I)} \big)^{- \mathsf{k}_I} 
 Q_{I-1}(S_s^{(I)})
 \label{CREapp}
\end{equation}
for $I=1,\cdots,M-1$. Here we understand that $Q_0(Z)=1$ and 
$Q_M(Z)=P_N(Z)$ where $P_N$ is defined in (\ref{PNZexp}).
Therefore, for $I=M-1$ the above expression gives the chiral ring equation (\ref{CR5dDualfinv1}) in the limit
when $\Lambda=0$. Using the explicit form of the functions $Q_I$ given in (\ref{QIZ}), we can rewrite
(\ref{CREapp}) as 
\begin{equation}
\begin{aligned}
\prod_{t=1}^{r_{I+1}} \big( S_s^{(I)} - S_t^{(I+1)}\big) &=
(-1)^{r_{I-1}} (\beta \Lambda_I )^{b_I}
 \big( S_s^{(I)}\big)^{\frac{b_I}{2}- \mathsf{k}_I}
 \\
 &\qquad\qquad\times\prod_{t=1}^{r_{I+1}} \big( S_t^{(I+1)}\big)^{\frac 12} 
  \prod_{u=1}^{r_{I-1}} \big( S_u^{(I-1)}\big)^{-\frac 12}
  \prod_{u=1}^{r_{I-1}} \big( S_s^{(I)} - S_u^{(I-1)}\big)
\end{aligned}
\label{CRgeneral}
\end{equation}
where we have used $b_I = r_{I+1}-r_{I-1}$. We now solve these equations for $S^{(I)}$
using the ansatz
\begin{equation}
S_{\star}^{(I)}=\diag\big(A_1,\, \dots \,,A_{r_{I-1}},\,A_{r_{I-1}+1} + \delta A_{r_{I-1}+1},\, \dots 
\, , A_{r_I}+\delta A_{r_I}\big)~.
\label{ansApp}
\end{equation}
Inserting this into (\ref{CRgeneral}), after some simple algebra we get
\begin{equation}
\delta A_s\,\prod_{\substack{t=1 \\ t \neq  s}}^{r_{I+1}} (A_s - A_t)
=(-1)^{r_{I-1}}(\beta\Lambda_I)^{b_I} (A_s)^{\frac{b_I}{2}- \mathsf{k}_I}
\prod_{t=1}^{r_{I+1}} ( A_t)^{\frac 12} 
  \prod_{u=1}^{r_{I-1}} (A_u)^{-\frac 12}
  \prod_{u=1}^{r_{I-1}} (A_s - A_u\big)
\label{chiral_ring_gen._op.2}
\end{equation}
for $s=r_{I-1}+1 ,\ldots ,\, r_I$. This leads to
\begin{equation}
\label{delta_As}
\delta A_s = (-1)^{r_{I-1}}(\beta\Lambda_I)^{b_I} 
(A_s)^{\frac{b_{I}}{2}-\mathsf{k}_I}
\!\!\!\!
\prod_{t=r_{I-1}+1}^{r_{I+1}} (A_t)^{\frac 12} 
\prod_{\substack{t=r_{I-1}+1\\ t \neq s}}^{r_{I+1}} \frac{1}{(A_s - A_t)}~.
\end{equation}
Using this in (\ref{ansApp}), we find that $\tr \log S_{\star}^{(I)}$ is a sum of $n_I$ terms, 
each of which looks very similar to the $q_I$-derivative of the localization result (\ref{WIgeneral}). 
Notice that the denominator of the latter is split into two products with $n_I-1$ and $n_{I+1}$ 
factors respectively, while the last product in (\ref{delta_As}) is written in terms of the ranks of 
the adjacent nodes and contains $r_{I+1} - r_{I-1} -1$ terms. However, using the relation between 
$n_I$ and $r_I$, we see that $r_{I+1} - r_{I-1}=n_{I} + n_{I+1}$, and thus the two structures agree. 
Actually, one can explicitly check that the chiral ring results fully match those from localization with the 
first JK prescription if 
\begin{equation}
q_I = -(-1)^{r_I} (\beta\Lambda_I)^{b_I} \quad\text{and}\quad
\mathsf{m}_I = \mathsf{k}_I
\end{equation}
for $I=1,\cdots,M-1$.

\subsubsection*{Dual quiver}

\begin{figure}[ht]
\centering
\begin{tikzpicture}[decoration={
markings,
mark=at position 0.6 with {\draw (-5pt,-5pt) -- (0pt,0pt);
                \draw (-5pt,5pt) -- (0pt,0pt);}}]
  \matrix[row sep=10mm,column sep=5mm] {
       \node(gfN)[flavor]{\Large $N$}; & & 
       \node(g2)[gauge] {\Large $\widetilde{r}_{1}$};  && 
       \node(dots){\Large $\ldots$}; & & 
       \node(glast)[gauge] {\Large $\widetilde{r}_{M-2}$};& &
      \node(g1)[gauge] {\Large $\widetilde{r}_{M-1}$}; \\
  };
\graph{(gfN) --[postaction={decorate}](g2) --[postaction={decorate}](dots)--[postaction={decorate}](glast) 
--[postaction={decorate}](g1);};
\end{tikzpicture}
\vspace{-0.5cm}
\caption{The quiver which is dual to the one in Fig.~\ref{3dquiverpicgeneric1}.}
\label{dualgeneric1}
\end{figure}
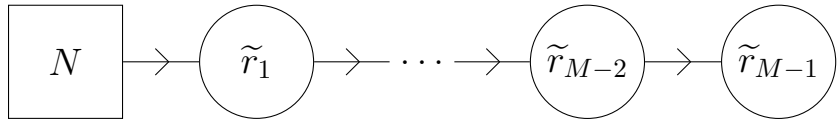

In a similar vein, we can treat the twisted chiral ring equations of the dual quiver which is represented
in Fig.~\ref{dualgeneric1}.

\noindent
In this case, the chiral ring equations take the form
\begin{equation}
\widetilde{Q}_{I-1}(\widetilde{S}^{(I)}_s)
= (-1)^{{\widetilde r}_{I-1}}\, (\beta\widetilde{\Lambda}_I)^{-\widetilde{b}_{I}}\, 
\big(\widetilde{S}^{(I)}_s\big)^{\widetilde{\mathsf{k}}_{I}} 
\widetilde{Q}_{I+1}(\widetilde{S}^{(I)}_s)
\end{equation}
for $I=1,\cdots,M-1$, where we understand that $\widetilde{Q}_0(Z) = P_N(Z)$ and $\widetilde{Q}_M(Z) 
= 1$. Again we notice that for $I=1$, the above formula reproduces the chiral ring equation
(\ref{dualCReq}) when $\Lambda=0$.

The analysis proceeds along the same lines as before. We first use the explicit expression of the
functions $\widetilde{Q}_I$ and get
\begin{align}
\label{TCRgeneral}
\prod_{t=1}^{\widetilde{r}_{I-1}} \big( \widetilde{S}_s^{(I)} - \widetilde{S}_t^{(I-1)}\big) &= 
(-1)^{\widetilde{r}_{I-1}} (\beta \widetilde{\Lambda}_I)^{-\widetilde{b}_I} 
\big( \widetilde{S}_s^{(I)}\big)^{\frac{b_I}{2}+\widetilde{\mathsf{k}}_I}\\
 &\qquad\qquad\times
 \prod_{t=1}^{\widetilde{r}_{I-1}} \big( \widetilde{S}_t^{(I-1)}\big)^{\frac 12}
 \prod_{u=1}^{\widetilde{r}_{I+1}} \big( \widetilde{S}_u^{(I+1)}\big)^{-\frac 12}
 \prod_{u=1}^{\widetilde{r}_{I+1}} \big( \widetilde{S}_s^{(I)} - \widetilde{S}_u^{(I+1)}\big)~.
 \notag
\end{align}
Next, using the fact that $\widetilde{r}_{I} = N - r_{I}$, we solve this equation 
for $\widetilde{S}_s^{(I)}$ with the ansatz
\begin{equation}
\label{SIdef}
\widetilde{S}^{(I)}_{\star} = \diag \big(
A_{r_{I}+1} + \delta A_{r_{I}+1}\,, \ldots
A_{r_{I+1}} + \delta A_{r_{I+1}}, A_{r_{I+1}+1}\, \ldots   A_{r_{I}+ \widetilde{r}_{I}} \big)
\end{equation}
and find
\begin{equation}
\begin{aligned}
\delta A_{r_{I}+s}&= (-1)^{\widetilde{r}_{I-1}}(\beta\widetilde{\Lambda}_I)^{-\widetilde{b}_I}
(A_{r_I +s})^{\frac{b_I}{2}+\widetilde{\mathsf{k}}_I}
\prod_{t=1}^{\widetilde{r}_{I-1}}(A_{r_{I-1}+t})^{\frac 12}
\prod_{u=1}^{\widetilde{r}_{I+1}}(A_{r_{I+1}+u})^{-\frac 12}\\
&\qquad\times
\prod_{u=1}^{\widetilde{r}_{I+1}}(A_{r_I +s}-A_{r_{I+1}+u})
\prod_{\stackrel{t=1}{t\ne s+n_I}}^{\widetilde{r}_{I-1}}\frac{1}{(A_{r_I +s}-A_{r_{I-1}+t})}~.
\end{aligned}
\end{equation}
There are lot of cancellations that take place between the products in the second line above, 
and in the end only $n_I +n_{I+1}-1$ of the terms survive as one can check by a careful analysis. 
Thus, we finally obtain
\begin{equation}
\delta A_{r_{I}+s}= (-1)^{\widetilde{r}_{I-1}}(\beta\widetilde{\Lambda}_I)^{-\widetilde{b}_I}
(A_{r_I +s})^{\frac{b_I}{2}+\widetilde{\mathsf{k}}_I}
\prod_{t=r_{I-1}+1}^{r_{I+1}} (A_t)^{\frac 12}
\prod_{\substack{t=r_{I-1}+1\\ t \neq s+n_I}}^{r_{I+1}} \frac{1}{( A_{r_I+s} - A_t)}~.
\end{equation}
Computing $\tr \log \widetilde{S}_{\star}^{(I)}$, we find it agrees with (negative of) the $q_I$-derivative 
of the twisted chiral superpotential (\ref{WIIgeneral}) obtained using the second JK prescription, 
provided we identify
\begin{equation}
q_I = -(-1)^{\widetilde{r}_{I}} (\beta\widetilde{\Lambda}_I)^{-\widetilde{b}_I} 
\quad\text{and}\quad \mathsf{m}_I = -\widetilde{\mathsf{k}}_I
\end{equation}
for $I=1,\cdots,M-1$.

This completes the identification of the parameters $\mathsf{m}_I$ with the Chern-Simons levels of the dual
pair of 3d quiver gauge theories studied in this work. 

\end{appendix}

\providecommand{\href}[2]{#2}\begingroup\raggedright\endgroup


\begin{thebibliography}{10}

\bibitem{Gukov:2006jk}
S.~Gukov and E.~Witten, {\it {Gauge Theory, Ramification, And The Geometric
  Langlands Program}},  \href{http://arxiv.org/abs/hep-th/0612073}{{\tt
  hep-th/0612073}}.

\bibitem{Gukov:2008sn}
S.~Gukov and E.~Witten, {\it {Rigid Surface Operators}},  {\em Adv. Theor.
  Math. Phys.} {\bf 14} (2010), no.~1 87--178,
  [\href{http://arxiv.org/abs/0804.1561}{{\tt arXiv:0804.1561}}].

\bibitem{Gukov:2014gja}
S.~Gukov, {\it {Surface Operators}},
  \href{http://arxiv.org/abs/1412.7127}{{\tt arXiv:1412.7127}}.

\bibitem{Gaiotto:2009fs}
D.~Gaiotto, {\it {Surface Operators in N = 2 4d Gauge Theories}},  {\em JHEP}
  {\bf 11} (2012) 090, [\href{http://arxiv.org/abs/0911.1316}{{\tt
  arXiv:0911.1316}}].

\bibitem{Alday:2009fs}
L.~F. Alday, D.~Gaiotto, S.~Gukov, Y.~Tachikawa, and H.~Verlinde, {\it {Loop
  and surface operators in N=2 gauge theory and Liouville modular geometry}},
  {\em JHEP} {\bf 1001} (2010) 113, [\href{http://arxiv.org/abs/0909.0945}{{\tt
  arXiv:0909.0945}}].

\bibitem{Taki:2009zd}
M.~Taki, {\it {On AGT Conjecture for Pure Super Yang-Mills and W-algebra}},
  {\em JHEP} {\bf 05} (2011) 038, [\href{http://arxiv.org/abs/0912.4789}{{\tt
  arXiv:0912.4789}}].

\bibitem{Alday:2010vg}
L.~F. Alday and Y.~Tachikawa, {\it {Affine SL(2) conformal blocks from 4d gauge
  theories}},  {\em Lett. Math. Phys.} {\bf 94} (2010) 87--114,
  [\href{http://arxiv.org/abs/1005.4469}{{\tt arXiv:1005.4469}}].

\bibitem{Kozcaz:2010yp}
C.~Kozcaz, S.~Pasquetti, F.~Passerini, and N.~Wyllard, {\it {Affine sl(N)
  conformal blocks from N=2 SU(N) gauge theories}},  {\em JHEP} {\bf 01} (2011)
  045, [\href{http://arxiv.org/abs/1008.1412}{{\tt arXiv:1008.1412}}].

\bibitem{Marshakov:2010fx}
A.~Marshakov, A.~Mironov, and A.~Morozov, {\it {On AGT Relations with Surface
  Operator Insertion and Stationary Limit of Beta-Ensembles}},  {\em
  J.Geom.Phys.} {\bf 61} (2011) 1203--1222,
  [\href{http://arxiv.org/abs/1011.4491}{{\tt arXiv:1011.4491}}].

\bibitem{Kozcaz:2010af}
C.~Kozcaz, S.~Pasquetti, and N.~Wyllard, {\it {A $\&$ B model approaches to
  surface operators and Toda theories}},  {\em JHEP} {\bf 08} (2010) 042,
  [\href{http://arxiv.org/abs/1004.2025}{{\tt arXiv:1004.2025}}].

\bibitem{Dimofte:2010tz}
T.~Dimofte, S.~Gukov, and L.~Hollands, {\it {Vortex Counting and Lagrangian
  3-manifolds}},  {\em Lett. Math. Phys.} {\bf 98} (2011) 225--287,
  [\href{http://arxiv.org/abs/1006.0977}{{\tt arXiv:1006.0977}}].

\bibitem{Maruyoshi:2010iu}
K.~Maruyoshi and M.~Taki, {\it {Deformed Prepotential, Quantum Integrable
  System and Liouville Field Theory}},  {\em Nucl. Phys.} {\bf B841} (2010)
  388--425, [\href{http://arxiv.org/abs/1006.4505}{{\tt arXiv:1006.4505}}].

\bibitem{Taki:2010bj}
M.~Taki, {\it {Surface Operator, Bubbling Calabi-Yau and AGT Relation}},  {\em
  JHEP} {\bf 07} (2011) 047, [\href{http://arxiv.org/abs/1007.2524}{{\tt
  arXiv:1007.2524}}].

\bibitem{Awata:2010bz}
H.~Awata, H.~Fuji, H.~Kanno, M.~Manabe, and Y.~Yamada, {\it {Localization with
  a Surface Operator, Irregular Conformal Blocks and Open Topological String}},
   {\em Adv. Theor. Math. Phys.} {\bf 16} (2012), no.~3 725--804,
  [\href{http://arxiv.org/abs/1008.0574}{{\tt arXiv:1008.0574}}].

\bibitem{Wyllard:2010vi}
N.~Wyllard, {\it {Instanton partition functions in N=2 SU(N) gauge theories
  with a general surface operator, and their W-algebra duals}},  {\em JHEP}
  {\bf 02} (2011) 114, [\href{http://arxiv.org/abs/1012.1355}{{\tt
  arXiv:1012.1355}}].

\bibitem{Wyllard:2010rp}
N.~Wyllard, {\it {W-algebras and surface operators in N=2 gauge theories}},
  {\em J. Phys.} {\bf A44} (2011) 155401,
  [\href{http://arxiv.org/abs/1011.0289}{{\tt arXiv:1011.0289}}].

\bibitem{Kanno:2011fw}
H.~Kanno and Y.~Tachikawa, {\it {Instanton counting with a surface operator and
  the chain-saw quiver}},  {\em JHEP} {\bf 06} (2011) 119,
  [\href{http://arxiv.org/abs/1105.0357}{{\tt arXiv:1105.0357}}].

\bibitem{Gaiotto:2013sma}
D.~Gaiotto, S.~Gukov, and N.~Seiberg, {\it {Surface Defects and Resolvents}},
  {\em JHEP} {\bf 09} (2013) 070, [\href{http://arxiv.org/abs/1307.2578}{{\tt
  arXiv:1307.2578}}].

\bibitem{Bullimore:2014awa}
M.~Bullimore, H.-C. Kim, and P.~Koroteev, {\it {Defects and Quantum
  Seiberg-Witten Geometry}},  {\em JHEP} {\bf 05} (2015) 095,
  [\href{http://arxiv.org/abs/1412.6081}{{\tt arXiv:1412.6081}}].

\bibitem{Nawata:2014nca}
S.~Nawata, {\it {Givental J-functions, Quantum integrable systems, AGT relation
  with surface operator}},  {\em Adv. Theor. Math. Phys.} {\bf 19} (2015)
  1277--1338, [\href{http://arxiv.org/abs/1408.4132}{{\tt arXiv:1408.4132}}].

\bibitem{Gomis:2014eya}
J.~Gomis and B.~Le~Floch, {\it {M2-brane surface operators and gauge theory
  dualities in Toda}},  {\em JHEP} {\bf 04} (2016) 183,
  [\href{http://arxiv.org/abs/1407.1852}{{\tt arXiv:1407.1852}}].

\bibitem{Frenkel:2015rda}
E.~Frenkel, S.~Gukov, and J.~Teschner, {\it {Surface Operators and Separation
  of Variables}},  {\em JHEP} {\bf 01} (2016) 179,
  [\href{http://arxiv.org/abs/1506.07508}{{\tt arXiv:1506.07508}}].

\bibitem{Assel:2016wcr}
B.~Assel and S.~Sch‰fer-Nameki, {\it {Six-dimensional origin of $ \mathcal{N} =
  4$ SYM with duality defects}},  {\em JHEP} {\bf 12} (2016) 058,
  [\href{http://arxiv.org/abs/1610.03663}{{\tt arXiv:1610.03663}}].

\bibitem{Gomis:2016ljm}
J.~Gomis, B.~Le~Floch, Y.~Pan, and W.~Peelaers, {\it {Intersecting Surface
  Defects and Two-Dimensional CFT}},  {\em Phys. Rev.} {\bf D96} (2017), no.~4
  045003, [\href{http://arxiv.org/abs/1610.03501}{{\tt arXiv:1610.03501}}].

\bibitem{Pan:2016fbl}
Y.~Pan and W.~Peelaers, {\it {Intersecting Surface Defects and Instanton
  Partition Functions}},  {\em JHEP} {\bf 07} (2017) 073,
  [\href{http://arxiv.org/abs/1612.04839}{{\tt arXiv:1612.04839}}].

\bibitem{Ashok:2017odt}
S.~K. Ashok, M.~Billo, E.~Dell'Aquila, M.~Frau, R.~R. John, and A.~Lerda, {\it
  {Modular and duality properties of surface operators in N=2* gauge
  theories}},  {\em JHEP} {\bf 07} (2017) 068,
  [\href{http://arxiv.org/abs/1702.02833}{{\tt arXiv:1702.02833}}].

\bibitem{Gorsky:2017hro}
A.~Gorsky, B.~Le~Floch, A.~Milekhin, and N.~Sopenko, {\it {Surface defects and
  instanton?vortex interaction}},  {\em Nucl. Phys.} {\bf B920} (2017)
  122--156, [\href{http://arxiv.org/abs/1702.03330}{{\tt arXiv:1702.03330}}].

\bibitem{Ashok:2017lko}
S.~K. Ashok, M.~Billo, E.~Dell'Aquila, M.~Frau, V.~Gupta, R.~R. John, and
  A.~Lerda, {\it {Surface operators, chiral rings, and localization in N=2
  gauge theories}},  {\em JHEP} {\bf 11} (2017) 137,
  [\href{http://arxiv.org/abs/1707.08922}{{\tt arXiv:1707.08922}}].

\bibitem{Nekrasov:2017rqy}
N.~Nekrasov, {\it {BPS/CFT correspondence IV: sigma models and defects in gauge
  theory}},  \href{http://arxiv.org/abs/1711.11011}{{\tt arXiv:1711.11011}}.

\bibitem{Nekrasov:2017gzb}
N.~Nekrasov, {\it {BPS/CFT correspondence V: BPZ and KZ equations from
  qq-characters}},  \href{http://arxiv.org/abs/1711.11582}{{\tt
  arXiv:1711.11582}}.

\bibitem{Nekrasov:2002qd}
N.~Nekrasov, {\it {Seiberg-Witten prepotential from instanton counting}},  {\em
  Adv. Theor. Math. Phys.} {\bf 7} (2004) 831--864,
  [\href{http://arxiv.org/abs/hep-th/0206161}{{\tt hep-th/0206161}}].

\bibitem{Nekrasov:2003rj}
N.~Nekrasov and A.~Okounkov, {\it {Seiberg-Witten theory and random
  partitions}},  {\em Prog. Math.} {\bf 244} (2006) 525--596,
  [\href{http://arxiv.org/abs/hep-th/0306238}{{\tt hep-th/0306238}}].

\bibitem{Seiberg:1994pq}
N.~Seiberg, {\it {Electric - magnetic duality in supersymmetric nonAbelian
  gauge theories}},  {\em Nucl. Phys.} {\bf B435} (1995) 129--146,
  [\href{http://arxiv.org/abs/hep-th/9411149}{{\tt hep-th/9411149}}].

\bibitem{Benini:2014mia}
F.~Benini, D.~S. Park, and P.~Zhao, {\it {Cluster Algebras from Dualities of 2d
  ${\mathcal{N}}$ = (2, 2) Quiver Gauge Theories}},  {\em Commun. Math. Phys.}
  {\bf 340} (2015) 47--104, [\href{http://arxiv.org/abs/1406.2699}{{\tt
  arXiv:1406.2699}}].

\bibitem{Closset:2015rna}
C.~Closset, S.~Cremonesi, and D.~S. Park, {\it {The equivariant A-twist and
  gauged linear sigma models on the two-sphere}},  {\em JHEP} {\bf 06} (2015)
  076, [\href{http://arxiv.org/abs/1504.06308}{{\tt arXiv:1504.06308}}].

\bibitem{Aharony:1997gp}
O.~Aharony, {\it {IR duality in d = 3 N=2 supersymmetric USp(2N(c)) and U(N(c))
  gauge theories}},  {\em Phys. Lett.} {\bf B404} (1997) 71--76,
  [\href{http://arxiv.org/abs/hep-th/9703215}{{\tt hep-th/9703215}}].

\bibitem{Aharony:2014uya}
O.~Aharony and D.~Fleischer, {\it {IR Dualities in General 3d Supersymmetric
  SU(N) QCD Theories}},  {\em JHEP} {\bf 02} (2015) 162,
  [\href{http://arxiv.org/abs/1411.5475}{{\tt arXiv:1411.5475}}].

\bibitem{Tachikawa:2004ur}
Y.~Tachikawa, {\it {Five-dimensional Chern-Simons terms and Nekrasov's
  instanton counting}},  {\em JHEP} {\bf 02} (2004) 050,
  [\href{http://arxiv.org/abs/hep-th/0401184}{{\tt hep-th/0401184}}].

\bibitem{Kim:2012gu}
H.-C. Kim, S.-S. Kim, and K.~Lee, {\it {5-dim Superconformal Index with
  Enhanced $E_n$ Global Symmetry}},  {\em JHEP} {\bf 10} (2012) 142,
  [\href{http://arxiv.org/abs/1206.6781}{{\tt arXiv:1206.6781}}].

\bibitem{Bergman:2013ala}
O.~Bergman, D.~RodrÌguez-GÛmez, and G.~Zafrir, {\it {Discrete $\theta$ and the
  5d superconformal index}},  {\em JHEP} {\bf 01} (2014) 079,
  [\href{http://arxiv.org/abs/1310.2150}{{\tt arXiv:1310.2150}}].

\bibitem{Bergman:2013aca}
O.~Bergman, D.~RodrÌguez-GÛmez, and G.~Zafrir, {\it {5-Brane Webs, Symmetry
  Enhancement, and Duality in 5d Supersymmetric Gauge Theory}},  {\em JHEP}
  {\bf 03} (2014) 112, [\href{http://arxiv.org/abs/1311.4199}{{\tt
  arXiv:1311.4199}}].

\bibitem{Taki:2013vka}
M.~Taki, {\it {Notes on Enhancement of Flavor Symmetry and 5d Superconformal
  Index}},  \href{http://arxiv.org/abs/1310.7509}{{\tt arXiv:1310.7509}}.

\bibitem{Taki:2014pba}
M.~Taki, {\it {Seiberg Duality, 5d SCFTs and Nekrasov Partition Functions}},
  \href{http://arxiv.org/abs/1401.7200}{{\tt arXiv:1401.7200}}.

\bibitem{Hwang:2014uwa}
C.~Hwang, J.~Kim, S.~Kim, and J.~Park, {\it {General instanton counting and 5d
  SCFT}},  {\em JHEP} {\bf 07} (2015) 063,
  [\href{http://arxiv.org/abs/1406.6793}{{\tt arXiv:1406.6793}}]. [Addendum:
  JHEP04,094(2016)].

\bibitem{JK1995}
L.~C. Jeffrey and F.~C. Kirwan, {\it {Surface Operators and Separation of
  Variables}},  {\em Topology} {\bf 34} (1995) 291--327.

\bibitem{Niemi:1983rq}
A.~J. Niemi and G.~W. Semenoff, {\it {Axial Anomaly Induced Fermion
  Fractionization and Effective Gauge Theory Actions in Odd Dimensional
  Space-Times}},  {\em Phys. Rev. Lett.} {\bf 51} (1983) 2077.

\bibitem{Redlich:1983kn}
A.~N. Redlich, {\it {Gauge Noninvariance and Parity Violation of
  Three-Dimensional Fermions}},  {\em Phys. Rev. Lett.} {\bf 52} (1984) 18.

\bibitem{Redlich:1983dv}
A.~N. Redlich, {\it {Parity Violation and Gauge Noninvariance of the Effective
  Gauge Field Action in Three-Dimensions}},  {\em Phys. Rev.} {\bf D29} (1984)
  2366--2374.

\bibitem{Benini:2011mf}
F.~Benini, C.~Closset, and S.~Cremonesi, {\it {Comments on 3d Seiberg-like
  dualities}},  {\em JHEP} {\bf 10} (2011) 075,
  [\href{http://arxiv.org/abs/1108.5373}{{\tt arXiv:1108.5373}}].

\bibitem{Intriligator:1997pq}
K.~A. Intriligator, D.~R. Morrison, and N.~Seiberg, {\it {Five-dimensional
  supersymmetric gauge theories and degenerations of Calabi-Yau spaces}},  {\em
  Nucl. Phys.} {\bf B497} (1997) 56--100,
  [\href{http://arxiv.org/abs/hep-th/9702198}{{\tt hep-th/9702198}}].

\bibitem{Jefferson:2017ahm}
P.~Jefferson, H.-C. Kim, C.~Vafa, and G.~Zafrir, {\it {Towards Classification
  of 5d SCFTs: Single Gauge Node}},
  \href{http://arxiv.org/abs/1705.05836}{{\tt arXiv:1705.05836}}.

\bibitem{Jefferson:2018irk}
P.~Jefferson, S.~Katz, H.-C. Kim, and C.~Vafa, {\it {On Geometric
  Classification of 5d SCFTs}},  \href{http://arxiv.org/abs/1801.04036}{{\tt
  arXiv:1801.04036}}.

\bibitem{Hollowood:2003cv}
T.~J. Hollowood, A.~Iqbal, and C.~Vafa, {\it {Matrix models, geometric
  engineering and elliptic genera}},  {\em JHEP} {\bf 03} (2008) 069,
  [\href{http://arxiv.org/abs/hep-th/0310272}{{\tt hep-th/0310272}}].

\bibitem{Billo:2012st}
M.~Billo, M.~Frau, F.~Fucito, L.~Giacone, A.~Lerda, J.~F. Morales, and
  D.~Ricci-Pacifici, {\it {Non-perturbative gauge/gravity correspondence in N=2
  theories}},  {\em JHEP} {\bf 1208} (2012) 166,
  [\href{http://arxiv.org/abs/1206.3914}{{\tt arXiv:1206.3914}}].

\bibitem{Seiberg:1994rs}
N.~Seiberg and E.~Witten, {\it {Monopole condensation, and confinement in N=2
  supersymmetric Yang-Mills theory}},  {\em Nucl. Phys.} {\bf B426} (1994)
  19--52, [\href{http://arxiv.org/abs/hep-th/9407087}{{\tt hep-th/9407087}}].

\bibitem{Nekrasov:1996cz}
N.~Nekrasov, {\it {Five dimensional gauge theories and relativistic integrable
  systems}},  {\em Nucl. Phys.} {\bf B531} (1998) 323--344,
  [\href{http://arxiv.org/abs/hep-th/9609219}{{\tt hep-th/9609219}}].

\bibitem{Hanany:2005hq}
A.~Hanany, P.~Kazakopoulos, and B.~Wecht, {\it {A New infinite class of quiver
  gauge theories}},  {\em JHEP} {\bf 08} (2005) 054,
  [\href{http://arxiv.org/abs/hep-th/0503177}{{\tt hep-th/0503177}}].

\bibitem{Brini:2008rh}
A.~Brini and A.~Tanzini, {\it {Exact results for topological strings on
  resolved $Y^{p,q}$ singularities}},  {\em Commun. Math. Phys.} {\bf 289}
  (2009) 205--252, [\href{http://arxiv.org/abs/0804.2598}{{\tt
  arXiv:0804.2598}}].

\bibitem{Aharony:1997ju}
O.~Aharony and A.~Hanany, {\it {Branes, superpotentials and superconformal
  fixed points}},  {\em Nucl. Phys.} {\bf B504} (1997) 239--271,
  [\href{http://arxiv.org/abs/hep-th/9704170}{{\tt hep-th/9704170}}].

\bibitem{Aharony:1997bh}
O.~Aharony, A.~Hanany, and B.~Kol, {\it {Webs of (p,q) five-branes,
  five-dimensional field theories and grid diagrams}},  {\em JHEP} {\bf 01}
  (1998) 002, [\href{http://arxiv.org/abs/hep-th/9710116}{{\tt
  hep-th/9710116}}].

\bibitem{Hori:2000kt}
K.~Hori and C.~Vafa, {\it {Mirror symmetry}},
  \href{http://arxiv.org/abs/hep-th/0002222}{{\tt hep-th/0002222}}.

\bibitem{Witten:1997sc}
E.~Witten, {\it {Solutions of four-dimensional field theories via M theory}},
  {\em Nucl.Phys.} {\bf B500} (1997) 3--42,
  [\href{http://arxiv.org/abs/hep-th/9703166}{{\tt hep-th/9703166}}].

\bibitem{Brandhuber:1997ua}
A.~Brandhuber, N.~Itzhaki, J.~Sonnenschein, S.~Theisen, and S.~Yankielowicz,
  {\it {On the M theory approach to (compactified) 5-D field theories}},  {\em
  Phys. Lett.} {\bf B415} (1997) 127--134,
  [\href{http://arxiv.org/abs/hep-th/9709010}{{\tt hep-th/9709010}}].

\bibitem{Wijnholt:2004rg}
M.~Wijnholt, {\it {Five-dimensional gauge theories and unitary matrix models}},
   \href{http://arxiv.org/abs/hep-th/0401025}{{\tt hep-th/0401025}}.

\bibitem{Cachazo:2002ry}
F.~Cachazo, M.~R. Douglas, N.~Seiberg, and E.~Witten, {\it {Chiral rings and
  anomalies in supersymmetric gauge theory}},  {\em JHEP} {\bf 12} (2002) 071,
  [\href{http://arxiv.org/abs/hep-th/0211170}{{\tt hep-th/0211170}}].

\bibitem{Kitao:1998mf}
T.~Kitao, K.~Ohta, and N.~Ohta, {\it {Three-dimensional gauge dynamics from
  brane configurations with (p,q) - five-brane}},  {\em Nucl. Phys.} {\bf B539}
  (1999) 79--106, [\href{http://arxiv.org/abs/hep-th/9808111}{{\tt
  hep-th/9808111}}].

\bibitem{Bergman:1999na}
O.~Bergman, A.~Hanany, A.~Karch, and B.~Kol, {\it {Branes and supersymmetry
  breaking in three-dimensional gauge theories}},  {\em JHEP} {\bf 10} (1999)
  036, [\href{http://arxiv.org/abs/hep-th/9908075}{{\tt hep-th/9908075}}].

\bibitem{Hanany:1997vm}
A.~Hanany and K.~Hori, {\it {Branes and N=2 theories in two-dimensions}},  {\em
  Nucl. Phys.} {\bf B513} (1998) 119--174,
  [\href{http://arxiv.org/abs/hep-th/9707192}{{\tt hep-th/9707192}}].

\bibitem{Chen:2012we}
H.-Y. Chen, T.~J. Hollowood, and P.~Zhao, {\it {A 5d/3d duality from
  relativistic integrable system}},  {\em JHEP} {\bf 07} (2012) 139,
  [\href{http://arxiv.org/abs/1205.4230}{{\tt arXiv:1205.4230}}].

\bibitem{Nekrasov:2009ui}
N.~A. Nekrasov and S.~L. Shatashvili, {\it {Quantum integrability and
  supersymmetric vacua}},  {\em Prog. Theor. Phys. Suppl.} {\bf 177} (2009)
  105--119, [\href{http://arxiv.org/abs/0901.4748}{{\tt arXiv:0901.4748}}].

\bibitem{Nekrasov:2009rc}
N.~Nekrasov and S.~Shatashvili, {\it {Quantization of Integrable Systems and
  Four Dimensional Gauge Theories}},
  \href{http://arxiv.org/abs/0908.4052}{{\tt arXiv:0908.4052}}.

\bibitem{Bruzzo:2002xf}
U.~Bruzzo, F.~Fucito, J.~F. Morales, and A.~Tanzini, {\it {Multi-instanton
  calculus and equivariant cohomology}},  {\em JHEP} {\bf 05} (2003) 054,
  [\href{http://arxiv.org/abs/hep-th/0211108}{{\tt hep-th/0211108}}].

\bibitem{Losev:2003py}
A.~S. Losev, A.~Marshakov, and N.~A. Nekrasov, {\it {Small instantons, little
  strings and free fermions}},  \href{http://arxiv.org/abs/hep-th/0302191}{{\tt
  hep-th/0302191}}.

\bibitem{Flume:2004rp}
R.~Flume, F.~Fucito, J.~F. Morales, and R.~Poghossian, {\it {Matone's relation
  in the presence of gravitational couplings}},  {\em JHEP} {\bf 04} (2004)
  008, [\href{http://arxiv.org/abs/hep-th/0403057}{{\tt hep-th/0403057}}].

\bibitem{Ashok:2016ewb}
S.~K. Ashok, M.~Billo, E.~Dell'Aquila, M.~Frau, A.~Lerda, M.~Moskovic, and
  M.~Raman, {\it {Chiral observables and S-duality in N = 2* U(N) gauge
  theories}},  {\em JHEP} {\bf 11} (2016) 020,
  [\href{http://arxiv.org/abs/1607.08327}{{\tt arXiv:1607.08327}}].

\end{thebibliography}

\end{document}